\pdfoutput=1
\documentclass[a4paper,12pt]{article}
\usepackage{jheppub}
\usepackage{amsmath}
\usepackage{amssymb}
\usepackage{braket}
\usepackage{cancel}
\usepackage{graphicx}
\usepackage[utf8]{inputenc}
\usepackage{hyperref}
\usepackage{multirow}
\usepackage{slashed}
\usepackage{booktabs}
\usepackage[normalem]{ulem}
\usepackage{wasysym}
\usepackage{braket}
\usepackage{simplewick}
\usepackage{array}
\usepackage{float}
\usepackage{titlesec}
\usepackage{listings}
\usepackage{bbold}
\usepackage[english]{babel}
\usepackage{subcaption}
\usepackage{braket}
\usepackage{comment}
\usepackage{enumitem}
\usepackage{bm}
\usepackage[export]{adjustbox}
\allowdisplaybreaks

\newcommand{\rd}{\mathrm{d}}
\newcommand{\re}{\mathrm{e}}
\newcommand{\ri}{\mathrm{i}}

\newcommand{\gammaE}{\gamma_{\mathrm{E}}}

\newcommand{\alphas}{\alpha_{\mathrm{s}}}
\newcommand{\alphaem}{\alpha_{\mathrm{em}}}
\newcommand{\GF}{G_{\mathrm{F}}}

\newcommand{\PR}{P_{\mathrm{R}}}

\newcommand{\CF}{C_\mathrm{F}}
\newcommand{\CA}{C_\mathrm{A}}
\newcommand{\TF}{T_\mathrm{F}}
\newcommand{\Nc}{N_\mathrm{c}}
\newcommand{\nl}{n_\mathrm{l}}
\newcommand{\nh}{n_\mathrm{h}}

\usepackage{lscape}
\usepackage[dvipsnames]{xcolor}

\begin{document}

\preprint{
\begin{flushright}
CERN-TH-2026-049,
P3H-26-018,
TTP26-007,
ZU-TH 11/26
\end{flushright}
}

\title{
\boldmath
The photon-energy spectrum
in $B\to X_s\gamma$ to N$^3$LO:
light-fermion and large-$\Nc$ corrections
}

\author[a,b]{Matteo Fael,}
\author[c,d]{Fabian Lange,}
\author[e]{Kay Sch\"onwald,}
\author[f]{and Matthias Steinhauser}

\affiliation[a]{
Dipartimento di Fisica e Astronomia ``G.~Galilei,'' Università di Padova,
\\
Via F.~Marzolo 8, 35131
Padova, Italy
}

\affiliation[b]{
Istituto Nazionale di Fisica Nucleare, Sezione di Padova,
\\ Via F.~Marzolo 8, 35131 Padova, Italy
}

\affiliation[c]{
Physik-Institut, Universität Zürich,\\
Winterthurerstrasse 190, 8057 Zürich, Switzerland}

\affiliation[d]{PSI Center for Neutron and Muon Sciences,\\
5232 Villigen PSI, Switzerland}

\affiliation[e]{
Theoretical Physics Department, CERN,\\
1211 Geneva, Switzerland}

\affiliation[f]{
Institut f\"ur Theoretische Teilchenphysik, Karlsruhe Institute of Technology (KIT),\\
Wolfgang-Gaede-Stra\ss{}e 1,
76131 Karlsruhe, Germany}

\emailAdd{matteo.fael@pd.infn.it}
\emailAdd{fabian.lange@physik.uzh.ch}
\emailAdd{kay.schonwald@cern.ch}
\emailAdd{matthias.steinhauser@kit.edu}

\abstract{
We calculate the photon-energy spectrum of the inclusive radiative decay
$B\to X_s\gamma$, induced by the electromagnetic dipole operator
$O_7$, to next-to-next-to-next-to-leading order
and consider the complete corrections for light fermions, for the contributions with two closed massive fermion loops, and for the limit of large QCD colour factors
$\Nc$ in the remaining part.
We discuss the total decay rate
both without and with a cut on the photon energy.
In addition to the on-shell renormalization of the bottom-quark mass, we also consider the kinetic mass and the MSR mass schemes.
The latter two lead to an improved perturbative behaviour of the decay rate.
}

\maketitle
\flushbottom

\newpage

\section{Introduction}
The inclusive radiative decay $B \to X_s \gamma$ is a cornerstone of flavour physics,
providing a high-precision test of the Standard Model (SM) and a powerful probe of potential
new-physics effects.
At the quark level, this process is mediated by the flavour-changing neutral-current (FCNC)
transition $b \to s \gamma$, which is absent at tree level in the SM due to the Glashow-Iliopoulos-Maiani (GIM) mechanism
and arises only through loop diagrams. As a result, the decay rate is strongly suppressed and
particularly sensitive to contributions from heavy virtual particles circulating in the loop.

From a theoretical point of view, the inclusive character of $B \to X_s \gamma$ permits a
systematic analysis within the framework of the operator product expansion combined with
perturbative quantum chromodynamics (QCD), leading to precise and reliable SM predictions. Nonperturbative effects
enter only at subleading order and are suppressed by powers of $\Lambda_{\mathrm{QCD}}/m_b$,
allowing them to be treated in a controlled manner.
Furthermore, in contrast to $b \to s \ell^+ \ell^-$ transitions, long-distance contributions
from intermediate resonant $c\bar c$ states are absent, rendering $B \to X_s \gamma$ one of
the theoretically cleanest FCNC observables.

On the experimental side, the branching fraction of $B \to X_s \gamma$
has been measured with steadily improving accuracy~\cite{CLEO:2001gsa,BaBar:2007yhb,Belle:2009nth,BaBar:2012fqh,BaBar:2012eja,Belle:2014nmp,HeavyFlavorAveragingGroupHFLAV:2024ctg}
and is in good agreement with SM predictions~\cite{Misiak:2015xwa,Czakon:2015exa,Misiak:2020vlo}.
The Belle II experiment is expected to reach a precision of roughly $2.5\%$
on the branching fraction~\cite{Belle-II:2022cgf}.
To match this level of experimental accuracy and to fully exploit the sensitivity of this
observable to new physics, further improvements in the theoretical description of
$B \to X_s \gamma$ are clearly required.

In this work, we concentrate on the photon-energy spectrum in $B \to X_s \gamma$.
Since experimental analyses impose a lower cut on the photon energy of about
$1.8$--$2.0\,\mathrm{GeV}$ in the $B$-meson rest frame, an accurate theoretical prediction
for the differential decay rate is essential for a meaningful comparison with data.
In this region of phase space, perturbative QCD corrections dominate and
must therefore be computed with high precision.

Within soft-collinear effective theory (SCET), the photon spectrum in the
endpoint region is governed by the leading shape function
\cite{Neubert:1993um,Bigi:1993ex,Neubert:1993ch},
a universal nonperturbative object that encapsulates the QCD dynamics close to
the kinematic boundary and plays a role similar to that of parton distribution functions.
The same shape function also appears in the theoretical description of inclusive semileptonic
decays $B \to X_u \ell \bar{\nu}_\ell$ and is thus a crucial ingredient in determinations of $|V_{ub}|$.
A robust extraction of the shape function from experimental data hinges on an accurate
control of the short-distance contributions to the spectrum, including higher-order perturbative effects.

The SIMBA collaboration has performed a fit to the $B \to X_s \gamma$ photon
spectrum using fixed-order theoretical predictions at next-to-next-to-leading order (NNLO),
next-to-next-to-leading-logarithmic (NNLL) resummation
plus all singular terms at $O(\alpha_s^2)$ (NNLL$^\prime$)~\cite{Ligeti:2008ac,Bernlochner:2020jlt}.
More recently, this framework was extended to next-to-next-to-next-to-leading logarithmic (N$^3$LL$^\prime$) accuracy in Ref.~\cite{Dehnadi:2022prz}.
However, that analysis did not yet include the three-loop hard function,
corresponding to the $b \to s \gamma$ form factor,
nor the complete fixed-order N$^3$LO result for the photon spectrum.
These missing ingredients were therefore treated as nuisance parameters,
and their variation induced an additional theory uncertainty.
The hard function at $O(\alphas^3)$ has since been computed in Ref.~\cite{Fael:2024vko}.

At the parton level, the photon-energy spectrum in $B \to X_s \gamma$ is known
up to NNLO for the contributions arising from the interference of
the electromagnetic and chromomagnetic dipole operators~\cite{Melnikov:2005bx,Asatrian:2006sm,Asatrian:2010rq}.
For the interference between four-quark
current-current
operators and the electromagnetic
dipole operator, only an interpolation between the heavy-charm limit~\cite{Misiak:2006ab,Misiak:2010sk}
and the massless-charm limit~\cite{Czakon:2015exa}
is currently available. Partial NNLO results for the two-body contribution
with exact charm-mass dependence have been reported recently~\cite{Greub:2023msv,Fael:2023gau,Czaja:2023ren,Greub:2024mwp}.
In addition, perturbative corrections to the matching~\cite{Misiak:2004ew}
and renormalization-group evolution~\cite{Gorbahn:2004my,Gorbahn:2005sa,Czakon:2006ss}
of the operators in the $\Delta B = \Delta S = 1$ effective Hamiltonian are known through NNLO.

In this paper, we take a first step towards a full N$^3$LO determination of the
photon-energy spectrum in QCD.
We compute the contribution of the electromagnetic dipole operator
$O_7$ to $\rd\Gamma/\rd E_\gamma$ through $O(\alphas^3)$ and
present analytic expressions for the fermionic colour structures proportional
to $\nl^2$, $\nl$, and $\nh^2$, as well as for the leading-colour
bosonic contribution proportional to $\Nc^3$ at N$^3$LO.
In particular, we evaluate both the singular and non-singular terms in the spectrum
arising from real-emission corrections.
By combining these results with the virtual contributions inferred
from the three-loop tensor form factors of Ref.~\cite{Fael:2024vko},
we also obtain the total decay rate through $O(\alphas^3)$.

The paper is organized as follows: In the next Section we introduce the relevant parts of the effective theory necessary
to obtain the photon-energy spectrum and explain
the strategy for the evaluation of virtual and real corrections.
The latter is the main new result obtained in this paper.
In Section~\ref{sec::details} we describe in detail the techniques, which we use to
extract the singular and regular part of the spectrum.
Our analytical and numerical results
are presented in Section~\ref{sec::res}
and the conclusions can be found in Section~\ref{sec::concl}. In the Appendix we provide useful formulae
in connection to the plus distributions and explicit analytic results for the
regular part of the photon spectrum.

\section{\label{sec::spectrum}Photon-energy spectrum to higher orders in QCD}

We consider the decay of a $B$ meson into a photon and any hadronic final state
containing a strange quark: $B \to X_s \gamma$. To leading order in the heavy
quark expansion (HQE) the decay is described by the decay $b \to X_s \gamma $ of a free bottom quark.
In the SM the decay is governed by the $\Delta B = \Delta S = 1$ effective Lagrangian
\begin{equation}
    \mathcal{L}_\mathrm{eff} =
    \frac{4 \GF}{\sqrt{2}}
    V_{tb} V_{ts}^\star
    \left[
    \sum_{i=1}^6
    C_i(\mu) O_i
    +C_7(\mu) O_7
    +C_8(\mu) O_8
    \right],
\end{equation}
where $O_{1} , \dots , O_{6}$ are the usual current-current and
QCD penguin operators, while $O_7$ and $O_8$ are
the electro- and chromo-magnetic dipole operators.
In this paper, we consider the contribution to the $B \to X_s \gamma$
rate originating only from the electromagnetic dipole operator
\begin{equation}
    O_7 =
    \frac{e}{16\pi^2} \overline{m}_b
    \Big( \bar s \sigma^{\mu\nu} \PR b \Big) F_{\mu\nu} .
\end{equation}
%\fl{which most likely provides the leading corrections.}
%\flnote{Reference or short, intuitive argument to support the statement?}
Here $F_{\mu\nu}$ is the electromagnetic field strength tensor, $e$ is the electromagnetic coupling constant with $\alphaem = e^2/(4\pi)$,
$\sigma_{\mu\nu} = \ri[\gamma_\mu,\gamma_\nu]/2$,
and the bottom quark mass $\overline{m}_b \equiv \overline{m}_b(\mu)$ in the $\overline{\mathrm{MS}}$ scheme.

The photon energy spectrum can be written in terms of the
normalized photon energy $x = 2 E_\gamma/m_b$ in the $b$ rest frame as
\begin{equation}
    \frac{\rd \Gamma}{\rd x} =
    \frac{\GF^2 \alphaem \overline{m}_b (\mu) m_b^3}{32 \pi ^4}
    |V_{tb} V_{ts}^\star|^2
    \sum_{i,j}
    C_i(\mu) C_j(\mu)
    \frac{\rd G_{ij}(x,\mu)}{\rd x},
    \label{eqn:dGammax}
\end{equation}
where $C_i(\mu)$ are the Wilson coefficients in the
effective theory evaluated at the scale $\mu \simeq m_b$.
We will focus on the function $\rd G_{77}/\rd x$ and compute it up to
N$^3$LO.
In the following we will denote the contribution of this term to the decay rate as $\Gamma_{77}$.
As a by-product, we will also obtain the prediction
for the total rate  by integrating Eq.~\eqref{eqn:dGammax} over
the whole phase space.
To higher orders in QCD, the spectrum $\rd \Gamma/\rd x$ receives
contributions from the two-body decay $b \to s \gamma$,
which we will simply refer to as ``virtual corrections'', as well as from
multi-parton emissions $b \to s \gamma g$, $b \to s \gamma g g$, etc.\
which we denote as ``real corrections.''

\subsubsection*{\label{sec::virt}Virtual corrections}
For the evaluation of the virtual corrections to the spectrum,
we make use of the results for the QCD heavy-to-light form factors
up to $O(\alphas^3)$
presented in Ref.~\cite{Fael:2024vko}.
In particular, we
concentrate on the tensor current
\begin{equation}
    j_t^{\mu\nu} =
    \ri \bar \psi_Q \sigma^{\mu\nu} \psi_q.
\end{equation}
Here we follow the notation of Ref.~\cite{Fael:2024vko} and define
the form factors as
\begin{align}
  \Gamma_{\mu\nu}^t(q_1,q_2) =&
  \ri F^t_1(q^2) \sigma_{\mu\nu}
  + \frac{F^t_2(q^2)}{m} \left( q_{1,\mu} \gamma_\nu - q_{1,\nu} \gamma_\mu \right)
  + \frac{F^t_3(q^2)}{m} \left( q_{2,\mu} \gamma_\nu - q_{2,\nu} \gamma_\mu \right)
  \nonumber\\ &
  + \frac{F^t_4(q^2)}{m^2} \left( q_{1,\mu}q_{2,\nu} - q_{1,\nu}q_{2,\mu} \right)
  \,.
  \label{eq::TensorFF}
\end{align}
The momentum $q_1$ is the incoming momentum of the massless quark and
$q_2$ is the outgoing momentum of the heavy quark. Furthermore, we
have $q=q_1-q_2$, with $q^2$, $q_1^2=0$ and $q_2^2=m^2$.

The virtual corrections to the differential rate can be obtained
from the form factors calculated at $q^2=0$. We obtain
the tree-level decay rate for $b \to s \gamma$ using as Feynman rule $ - 2 \ri \PR \Gamma_{\mu\nu}^t p_\gamma^\mu$, where we identify $q_1 = -p_s$, $q_2 = -p_b$, and $q = p_\gamma$.
The expression for the virtual corrections in terms of the
tensor form factor  is given by
\begin{equation}
\frac{\rd G_{77}^\mathrm{virt}}{\rd  x}
=
%\left( \frac{4\pi \mu^2}{m^2} \right)^\epsilon
 \frac{1 - \epsilon }{4}
\frac{\Gamma(1-\epsilon)\re^{\epsilon \gammaE}}{\Gamma(2-2\epsilon)}
 \Big[ 2 F^{t}_1(0) - F^{t}_2(0) - F^{t}_3(0) \Big]^2
\delta(1 - x).
\end{equation}

\subsubsection*{Real corrections}
\begin{figure}[t]
    \centering
    \begin{subfigure}{0.32\textwidth}
    \includegraphics[width=\textwidth]{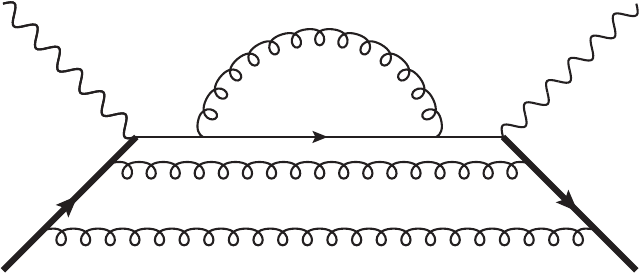}
    \caption{}
    \end{subfigure}
    \begin{subfigure}{0.32\textwidth}
    \includegraphics[width=\textwidth]{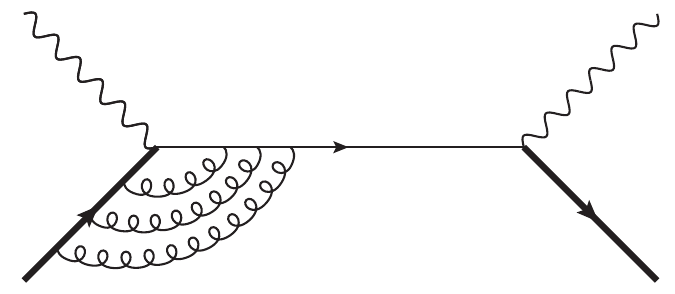}
    \caption{}
    \end{subfigure}
    \begin{subfigure}{0.32\textwidth}
    \includegraphics[width=\textwidth]{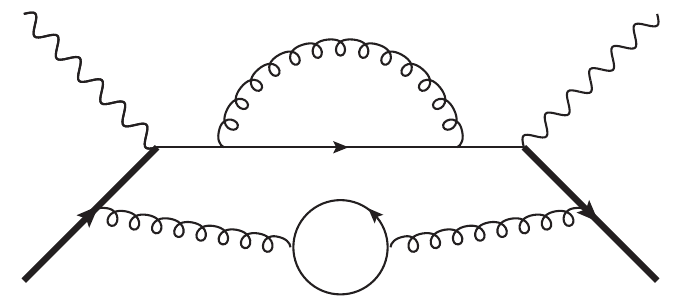}
    \caption{}
    \end{subfigure}
    \caption{Sample Feynman diagrams contributing to the forward scattering amplitude of $b \gamma \to b \gamma$. Their imaginary parts provide the real
    emission contribution to the photon energy spectrum in $B \to X_s \gamma$.
    Thin (bold) lines denote the strange (bottom) quarks.}
    \label{fig:fdreal}
\end{figure}
To compute the real-emission contributions with the emission of $n$ additional partons, we consider
forward scattering amplitudes for the process $b \gamma \to b \gamma$ and extract
their imaginary part (see the sample diagrams in Fig.~\ref{fig:fdreal}).

The differential decay rate of $b \to X_s \gamma$ can be written as
\begin{equation}
\rd  \Gamma =
\frac{(2\pi)^d}{2m_b} \,
\overline{|\mathcal{M}|^2} \,
\rd \Phi_{n+2}(p_b;  p_\gamma,p_s, p_1,\dots,p_n) ,
\end{equation}
where $p_\gamma$ and $p_s$ denote the four-momenta of the photon and strange quark in the rest frame of the $b$ quark, respectively,
and $d=4-2\epsilon$ is the space-time dimension.
The momenta of the additional real partons in the hadronic final state $X_s$ are denoted by $p_1, \dots, p_n$.
The quantity $\overline{|\mathcal{M}|^2}$ represents the squared matrix element, averaged over the spins of the initial state and summed over those of the final state.

A generic $n$-body phase space is defined as
\begin{equation}
    \rd \Phi_{n}(P; p_1, \dots, p_n) =
    \delta^{(d)}\left(P-\sum_{i=1}^n p_i\right) \prod_{i=1}^n
    \frac{\rd^{d-1} p_i}{(2\pi)^{d-1} 2E_i}.
\end{equation}
At leading order, the decay $b \to X_s \gamma$ is described by the two-body phase
space $\mathrm{d}\Phi_2$.
At higher orders in QCD, where $n$ additional partons are emitted, the
$(n+2)$-particle phase space can be written as
\begin{align}
    \rd \Phi_{n+2} &=
    \frac{\rd^{{d-1}}p_\gamma}{(2\pi)^{d-1} 2E_\gamma}
    \frac{\rd^{{d-1}}p_s}{(2\pi)^{d-1} 2E_s}
    \prod_{i=1}^n
    \frac{\rd^{{d-1}}p_i}{(2\pi)^{d-1} 2E_i}
    \delta^{(d)} \Big(p_b-p_\gamma-p_s-\sum_{i=1}^n p_i \Big)
    \notag \\
    & =
    \frac{E_\gamma^{d-3} \rd E_\gamma \rd \Omega_{d-1}}{2(2\pi)^{d-1}}
    \frac{\rd^{d-1} p_s}{(2\pi)^{d-1} 2E_s}
    \prod_{i=1}^n
    \frac{\rd^{{d-1}}p_i}{(2\pi)^{d-1} 2E_i}
    \delta^{(d)} \Big(p_b-p_\gamma-p_s-\sum_{i=1}^n p_i \Big).
\end{align}
After integrating over the $d$-dimensional solid angle of the photon momentum,
\begin{equation}
    \int \rd \Omega_{d-1} =
    (4 \pi)^{1-\epsilon}
    \frac{\Gamma(1-\epsilon)}{\Gamma(2-2\epsilon)},
\end{equation}
the differential decay rate with respect to the photon energy $E_\gamma$ can be
cast into the form
\begin{align}
    \frac{1}{E_\gamma^{1-2\epsilon}} \frac{\rd \Gamma}{\rd E_\gamma} &=
    \frac{1}{8 m \pi^{2-\epsilon}}
    \frac{\Gamma(1-\epsilon)}{\Gamma(2-2\epsilon)}
    \notag \\ & \quad
    \times (2\pi)^d \int
    \overline{|\mathcal{M}|^2} \,
    \frac{\rd^{d-1} p_s}{ 2E_s(2\pi)^{d-1}}
    \prod_{i=1}^n
    \frac{\rd^{{d-1}}p_i}{(2\pi)^{d-1} 2E_i}
    \delta^d\Big(p_b-p_\gamma-p_s-\sum_{i=1}^n p_i \Big).
    \label{eqn:dGammadEintegral}
\end{align}
If we perform the replacement $p_\gamma \to -p_\gamma$,
thereby interpreting the outgoing photon as an incoming one, the right-hand side of Eq.~\eqref{eqn:dGammadEintegral}---namely the factor isolated on the second line---can be identified with the total cross section for the scattering process $b \gamma \to X_s$ at fixed centre-of-mass energy
$(p_b + p_\gamma)^2 > m_b^2$.
This total cross section can be computed using the optical theorem, by evaluating the imaginary part of the forward scattering amplitude for $b \gamma \to b \gamma$.

\section{\label{sec::details}Technical Details}

\subsubsection*{Generation of the amplitude}
We generate the forward scattering diagrams with \texttt{qgraf}~\cite{Nogueira:1991ex} and
generate \texttt{FORM}~\cite{Ruijl:2017dtg,Davies:2026cci} code for the calculation of the amplitude with \texttt{tapir}~\cite{Gerlach:2022qnc} and \texttt{exp}~\cite{Harlander:1998cmq,Seidensticker:1999bb}.
\texttt{tapir} also generates partial fraction identities for the denominators, which are necessary due to the forward scattering kinematics.
The Dirac and colour algebra~\cite{vanRitbergen:1998pn} is performed with \texttt{FORM}.
Due to the partial fraction decomposition we end up with a large number of integral families which are not all independent.
We use \texttt{feynson}~\cite{Maheria:2022dsq} to map them to a minimal set.
In total we find 1, 6, and 44 integral families at one-, two- and three-loop order, respectively.

\subsubsection*{IBP reduction}
At one- and two-loop order, we perform a full integration-by-parts (IBP) reduction~\cite{Tkachov:1981wb,Chetyrkin:1981qh,Laporta:2000dsw}
of the scalar integrals with \texttt{Kira}~\cite{Maierhofer:2017gsa,Klappert:2020nbg,Lange:2025fba}.
We find 3 and 42 master integrals, respectively.
At two-loop order, 18 out of the 42 master integrals do not have an imaginary part and can therefore be discarded.

At three loops, the approach of first performing the full reduction to master integrals and discarding master integrals without imaginary part at a later stage leads to prohibitively expensive IBP reductions.
Therefore, we refine our approach in the following way.
Before reduction, we determine every sector of a family which does not generate an imaginary part and supply this sector as vanishing sector to \texttt{Kira}.
This way, integrals without imaginary part are dropped in every step, significantly reducing the complexity of the reduction (see also Ref.~\cite{Fael:2023tcv}).
Furthermore, the number of master integrals reduces drastically.
The use of \texttt{Kira~3}~\cite{Driesse:2024xad,Lange:2025fba} with \texttt{FireFly}~\cite{Klappert:2019emp,Klappert:2020aqs} as backend for finite field techniques~\cite{Kauers:2008zz,vonManteuffel:2014ixa,Peraro:2016wsq} was essential for performing the IBP reductions at three-loop order.
In order to facilitate the efficient reduction of the master integrals, we also search for a basis of master integrals, in which the denominators factorize in $d$ and $x$~\cite{Smirnov:2020quc,Usovitsch:2020jrk}.
To find suitable candidates for this search, we perform simpler test reductions with \texttt{Kira} and use \texttt{ImproveMasters.m}~\cite{Smirnov:2020quc}\footnote{It was shipped together with \texttt{FIRE} from version \texttt{6.4.1} until version \texttt{6.6}~\cite{Smirnov:2023yhb}.} to find a suitable basis.
In total we find 987 master integrals for the full amplitude.
Restricting ourselves to the large-$\Nc$ limit, only 132 master integrals are needed.
For the fermionic pieces with at least one massless fermion loop, we need 82 master integrals, and for the $\nh^2$ piece only 2 master integrals contribute.
Eliminating the overlap between these three contributions results in 158 master integrals which have to be computed.

\subsubsection*{Calculation of the master integrals}

To compute the master integrals, we derive differential equations in $x$~\cite{Kotikov:1990kg,Kotikov:1991hm,Kotikov:1991pm,Remiddi:1997ny}.
We employ \texttt{LiteRed}~\cite{Lee:2012cn,Lee:2013mka} to differentiate the master integrals with respect to $x$ and use reduction tables obtained by \texttt{Kira} to establish the differential equations.

The calculation of the master integrals is complicated by the soft singularity in the limit $y = 1-x \to 0$.
In this limit the master integrals behave as
\begin{align}
        \vec{I}(\epsilon,y) &= \sum\limits_{n} y^{m + n \epsilon} \vec{J}(\epsilon,y)~,
        \label{eq::branches}
\end{align}
which leads to distributions via the identities given in Appendix~\ref{app:distr}.
Therefore, we cannot simply calculate the master integrals as function of
$y$, but need to separate them into the different branches of the asymptotic expansion around $y \to 0$.
In general, we find branches with $n=1,2,...,2l$, where $l$ is the loop order under consideration.
Note that we do not have to consider the region with $n=0$, since this region does not contribute to the imaginary part of the integral.

To solve the integrals, we do the following:
\begin{itemize}
    \item We solve the integrals without separation into the different branches.
    To do this we use the algorithm outlined in Ref.~\cite{Ablinger:2018zwz} to solve the coupled system of first order differential equations in the variable $y=1-x$, i.e.\ we do not search for a canonical form of the differential equation, but solve the differential equations by decoupling to higher-order differential operators and factorization.
    This provides us with an analytic solution in terms of iterated integrals.
    At one- and two-loop order it is quite straightforward to obtain boundary conditions for $y\to0$ via the method of regions~\cite{Beneke:1997zp}, direct integration, and Mellin-Barnes methods,
    see e.g. Refs.~\cite{Smirnov:2012gma,Dubovyk:2022obc}.

    At three-loop order we proceed in a slightly different way. We use \texttt{AMFlow}~\cite{Liu:2017jxz,Liu:2021wks,Liu:2022mfb,Liu:2022chg} to obtain initial conditions at the point $y=2/11$ with 100 digits of precision.
    We compare these results with the numerical evaluation of the formal solutions of the master integrals in terms of iterated integrals, which still depend on undetermined boundary coefficients.
    For the numerical evaluation of the iterated integrals at $y=2/11$, we can perform a deep series expansion up to $\mathcal{O}(y^{200})$.
    We observe that this is sufficient to obtain 100 significant digits, and we can directly compare to the numerical evaluations with \texttt{AMFlow}.
    We found this approach to be more convenient, since for $y>0$ functions beyond multiple polylogarithms contribute to the results of the master integrals.
    In the end, we are able to obtain analytic  boundary conditions using the \texttt{PSLQ} algorithm~\cite{pslq}, and we find that only (multiple) $\zeta$ values appear.

    \item In a second step, we insert for each branch of the asymptotic expansion an ansatz of the form in Eq.~\eqref{eq::branches},
    where $\vec{J}(\epsilon,y)$ has a Taylor expansion around $y=0$,
    into the differential equation and express the result of each branch in terms of a minimal number of boundary conditions.
    We choose as starting power $m=-6$ such that $J$ has indeed a Taylor and not a Laurent expansion.
    By demanding that the sum of all branches recovers the expansion of the full master integrals we calculated in the step before, we can determine the majority of boundary conditions of the different branches.
    For some master integrals we are not able to fix all boundary conditions in this way.
    In these cases we either calculate the boundary conditions directly with the help of the method of regions, use consistency conditions at $y \to 1$, or expand the master integrals to higher orders in $\epsilon$.

    \item Having at our disposal the analytic expansion of each master integral around $y=0$ separated into branches, we solve the full differential equations again by using as boundary conditions these results for each branch individually.
    In this way we can express all master integrals as sums over all branches.
\end{itemize}

At one- and two-loop order we see that multiple polylogarithms of argument $y=1-x$ are enough to solve the differential equations.
The letters are given by:
\begin{align}
    f_{0}(t) &= \frac{1}{t}, &
    f_{-1}(t) &= \frac{1}{1+t}, &
    f_{1}(t) &= \frac{1}{1-t}, &
    f_{2}(t) &= \frac{1}{2-t}~.
\end{align}
At the three-loop level we find additionally
\begin{align}
    f_{a}(t) &= \sqrt{4-t}\sqrt{t}, &
    f_{b}(t) &= \frac{\sqrt{4-t}\sqrt{t}}{1-t}~.
    \label{eq::sqrt}
\end{align}
We use the definition of the iterated integrals
\begin{align}
    H_{w_1,\vec{w}}(x)
    &=
    \int\limits_{0}^{x}
    \mathrm{d} t
    f_{w_1} (t)
    H_{\vec{w}}(t)~,
    \label{eq::ints}
\end{align}
with the  usual regularizations for the letter $f_0(t)$, i.e.
\begin{align}
    H_{\underbrace{0,...,0}_{\text{n times}}}(x) &= \frac{1}{n!} \log^n(x)~.
\end{align}
It is interesting to note that the master integrals for the leading-$\Nc$ term do depend on the new letters, the final amplitude, however, is free of them.
The full complexity is only reached for the colour factors linear in $\nl$.
The colour factors which are not covered here also depend on more complicated geometries, like elliptic functions and beyond.

Note that it is in principle possible to rationalize the letters containing square-roots by a variable change of the argument, e.g.\ by changing to the variable $\omega = -(1-y)^2/y$.
This change of variables enables the numerical evaluation with established tools like \texttt{ginac}~\cite{Bauer:2000cp,Vollinga:2004sn},
but introduces many more letters and leads to a much bigger expression of the amplitude.
For an efficient numerical evaluation, we therefore expand the amplitude around $x=0$ and $x=1$ to 100 terms in each limit.
Both expansions have a radius of convergence of radius $1$ and agree to 30 digits at the intermediate point $x=1/2$.
The expansion around $x=0$ requires an analytic continuation, which we carry out with \texttt{HarmonicSums}~\cite{Vermaseren:1998uu,Blumlein:1998if,Blumlein:2009ta,Ablinger:2009ovq,Ablinger:2011te,Ablinger:2012ufz,Ablinger:2013eba,Ablinger:2013cf,Ablinger:2014bra,Ablinger:2014rba,Ablinger:2015gdg,Ablinger:2018cja}
and \texttt{Sigma}~\cite{sigmaI,sigmaII}.
Here, we encounter the constant
\begin{align}
    c_{1} &=
    H_{a}(1) H_{a,2,1}(1)
    - H_{a,a,2,1}(1)
    = 0.2598851170\ldots\,.
    \label{eq::c1}
\end{align}
For the total decay rate in Section~\ref{sub::total} we also encounter the constants
\begin{align}
    c_{2} &=
    H_{a}(1) H_{a,2,0}(1)
    - H_{a,a,2,0}(1)
    =
    -0.4018313022\ldots
    \nonumber\\
    c_{3} &=
    H_{a}(1) H_{a,2,1,0}(1)
    - H_{a,a,2,1,0}(1)
    =
    -0.3288306064\ldots
    \,.
    \label{eq::c23}
\end{align}
In the ancillary files~\cite{progdata} we provide
numerical evaluations of these constants with 1000 significant digits.
Expressing these constants in terms of Goncharov polylogarithms does not lead to a more compact notation.
However, if the letter $f_{2}(t)$ is absent, the rationalization leads only to harmonic and cyclotomic harmonic polylogarithms~\cite{Remiddi:1999ew,Ablinger:2011te} for which the transcendental constants at unit argument are known~\cite{Blumlein:2009cf,Ablinger:2011te}.
We have checked, using the \texttt{PSLQ} algorithm, that the constants in Eqs.~(\ref{eq::c1}) and~(\ref{eq::c23}) are linearly independent to these ones.

\subsubsection*{\label{sub::spectrum_reg_sing}Singular and regular  parts of the amplitude}

In order to separate the singular and regular parts
in the amplitude we proceed as follows. In a first step,
we insert the master integrals separated into the
different branches as described in the previous Section
into the amplitude. Next, we expand the
amplitude for $y\to0$ and
collect all contributions that behave as $1/y^{(k-n\epsilon)}$ with $k>0$. In our case, we have $k\le3$ and  $1\le n \le 6$.
These contributions define the singular part
of the amplitude, which we denote by
\begin{eqnarray}
    \frac{1}{E_\gamma} \frac{\rd G_{77}}{\rd  E_\gamma}\Bigg|_{\rm sing}
    \,.
\end{eqnarray}
The regular part is then obtained via
\begin{eqnarray}
    \frac{1}{E_\gamma} \frac{\rd G_{77}}{\rd  E_\gamma}\Bigg|_{\rm reg}
    &=&
 \frac{1}{E_\gamma} \frac{\rd G_{77}}{\rd    E_\gamma}
    -
    \frac{1}{E_\gamma} \frac{\rd G_{77}}{\rd  E_\gamma}\Bigg|_{\rm sing}
        ,
\end{eqnarray}
which has by construction a regular behaviour for $y\to0$.
Furthermore, it is finite for $\epsilon\to 0$ after renormalization.

At this point we use the relations
from Appendix~\ref{app:distr} to rewrite the
singular part in terms of distributions.
Afterwards, the coefficients of
the plus distributions are finite and the
limit $\epsilon\to0$ can be taken.
The coefficients of $\delta(y)$ still contain poles in $\epsilon$
up to $1/\epsilon^6$.
They cancel against the virtual corrections discussed in the previous Section.

\section{\label{sec::res}Results}

\subsection{Analytic expressions for the spectrum}

In the following we present the analytic results of our calculation for $\mu=m_b$.
The differential decay rate can be written as
\begin{align}
    \frac{\rd \Gamma_{77}}{\rd x} &=
    \Gamma_0
    \, m_b^3  \, \overline{m}_b^2(m_b)
    \, |C_7(m_b)|^2
    \Bigg[
    \delta(1-x)
    +\frac{\alphas}{\pi} Y_1(x)
    +\left(\frac{\alphas}{\pi} \right)^2
    Y_2(x)
    +\left(\frac{\alphas}{\pi} \right)^3
    Y_3(x)
    \Bigg]\,,
    \label{eq::spec}
\end{align}
where the expansion coefficients have the form
\begin{align}
    Y_{i}(x) &= Y_{i}^{\delta} \ \delta(1-x)  + Y_{i}^{+}(x) + Y_{i}^{\text{reg}}(x) ~.
\end{align}
The first two terms correspond to distributions, while the
last term is the regular part of the spectrum.
At NLO, NNLO, and N$^3$LO the terms proportional to the $\delta$-distribution are given by
\begin{small}
\begin{align}
    Y_1^{\delta} &=
    - \CF \biggl( \frac{5}{4} + \frac{\pi ^2}{3} \biggr)
    ~,\\
    Y_2^{\delta} &=
    \CA \CF
    \left(
        \frac{55 \zeta _3}{48}
        +\frac{143 \pi^4}{2880}
        -\frac{215 \pi^2}{864}
        -\frac{11987}{1728}
        -\frac{37}{24} \pi ^2 \ln(2)
    \right)
    \nonumber \\ &
    +\CF^2
    \left(
        -\frac{33 \zeta _3}{8}
        +\frac{59 \pi^4}{1440}
        -\frac{167 \pi ^2}{96}
        +\frac{989}{192}
        +\frac{37}{12}\pi ^2 \ln(2)
    \right)
    \nonumber \\ &
    +\CF \TF \nh
    \left(
        -\frac{\zeta _3}{3}
        -\frac{29 \pi^2}{54}
        +\frac{3563}{648}
    \right)
    + \CF \TF \nl
    \left(
        \frac{\zeta_3}{3}
        +\frac{91 \pi ^2}{216}
        +\frac{631}{432}
    \right)
    ~,\\
    Y_3^{\delta} &=
    \CA \CF \TF \nl
    \biggl(
        -\frac{1}{72} \pi ^2 \zeta _3
        +\frac{221\zeta _3}{432}
        -\frac{43 \zeta _5}{12}
        -\frac{67\text{Li}_4\left(\frac{1}{2}\right)}{9}
        +\frac{19 \pi^4}{648}
        +\frac{4001 \pi^2}{1944}
        +\frac{320753}{23328}
        \nonumber \\ &
        -\frac{67 \log^4(2)}{216}
        -\frac{23}{54} \pi ^2 \ln^2(2)
        +\frac{59}{18} \pi^2 \ln(2)
    \biggr)
    +\CF \TF^2 \nh \nl
    \biggl(
        \frac{4 \zeta_3}{27}
        -\frac{\pi ^4}{405}
        +\frac{443 \pi^2}{972}
        \nonumber \\ &
        -\frac{13243}{2916}
   \biggr)
   +\CF \TF^2 \nh^2
   \biggl(
        \frac{26\zeta _3}{9}
        -\frac{\pi ^4}{270}
        +\frac{2 \pi^2}{15}
        -\frac{374}{81}
    \biggr)
    +\CF \TF^2 \nl^2
    \biggl(
        -\frac{32 \zeta_3}{27}
        -\frac{\pi ^4}{108}
        \nonumber \\ &
        -\frac{1061 \pi^2}{1944}
        -\frac{12727}{11664}
    \biggr)
    +\CF^2 \TF \nl
    \biggl(
        \frac{5 \pi ^2 \zeta _3}{18}
        +\frac{1073 \zeta_3}{72}
        +\frac{10 \zeta _5}{3}
        +\frac{134\text{Li}_4\left(\frac{1}{2}\right)}{9}
        -\frac{6047 \pi^4}{12960}
        \nonumber \\ &
        +\frac{1181 \pi^2}{324}
        -\frac{13141}{2592}
        +\frac{67 \ln^4(2)}{108}
        +\frac{23}{27} \pi ^2 \ln^2(2)
        -\frac{59}{9} \pi^2 \ln(2)
    \biggr)
    + \Nc^3
    \biggl(
        -\frac{19 \zeta _3^2}{64}
        \nonumber \\ &
        +\frac{41\pi ^2 \zeta _3}{192}
        -\frac{16391 \zeta _3}{3456}
        +\frac{113\zeta _5}{48}
        -\frac{2455 \pi ^6}{290304}
        +\frac{111289 \pi^4}{414720}
        -\frac{277753 \pi^2}{124416}
        -\frac{11206753}{1492992}
    \biggr)
    \nonumber \\ &
    + \mathcal{O}(\Nc^2,\nh)
    ~.
    \label{eq::spec_delta}
\end{align}
\end{small}
The plus distributions read
\begin{small}
\begin{align}
    Y_1^{+}(x) &= -\CF \left(\frac{7 \mathcal{D}_0}{4} + \mathcal{D}_1\right)
    ~,\\
    Y_2^{+}(x) &=
    \CA \CF
    \biggl(
        \mathcal{D}_0
        \biggl[
            \frac{\zeta _3}{4}
            +\frac{17 \pi^2}{72}
            -\frac{905}{288}
        \biggr]
        +\mathcal{D}_1
        \biggl[
            \frac{95}{144}
            +\frac{\pi^2}{12}
        \biggr]
        + \frac{11 \mathcal{D}_2}{8}
    \biggr)
    \nonumber \\ &
    +\CF^2
    \biggl(
        \mathcal{D}_0
        \biggl[
            -\frac{\zeta _3}{2}
            +\frac{5 \pi^2}{12}
            +\frac{67}{32}
        \biggr]
        + \mathcal{D}_1
        \biggl[
            \frac{69}{16}
            +\frac{\pi^2}{6}
        \biggr]
        + \frac{21 \mathcal{D}_2}{8}
        + \frac{ \mathcal{D}_3}{2}
    \biggr)
    \nonumber \\ &
    + \CF \TF \nl
    \biggl(
        \mathcal{D}_0
        \biggl[
            \frac{85}{72}
            -\frac{\pi^2}{18}
        \biggr]
        -\frac{13 \mathcal{D}_1}{36}
        -\frac{\mathcal{D}_2}{2}
    \biggr)
    ~,\\
    Y_3^{+}(x) &=
    \CA \CF \TF \nl
    \biggl(
        \mathcal{D}_0
        \biggl[
            -\frac{11 \zeta_3}{18}
            +\frac{19 \pi ^4}{720}
            -\frac{619 \pi^2}{648}
            +\frac{4205}{648}
        \biggr]
        +\mathcal{D}_1
        \biggl[
            -\frac{\zeta_3}{6}
            +\frac{2 \pi^2}{9}
            -\frac{5399}{1296}
        \biggr]
        \nonumber \\ &
        +\mathcal{D}_2
        \biggl[
            \frac{\pi ^2}{12}
            -\frac{10}{9}
        \biggr]
        + \frac{77}{54} \mathcal{D}_3
    \biggr)
    +\CF^2 \TF \nl
    \biggl(
        \mathcal{D}_1
        \biggl[
            \frac{2 \zeta_3}{3}
            -\frac{47 \pi ^2}{108}
            -\frac{1025}{216}
        \biggr]
        +\mathcal{D}_0
        \biggl[
            \frac{37 \zeta _3}{18}
            +\frac{\pi ^4}{1080}
            \nonumber \\ &
            -\frac{1063\pi ^2}{864}
            -\frac{727}{432}
        \biggr]
        -\frac{55}{96} \mathcal{D}_2
        +\frac{55}{36} \mathcal{D}_3
        +\frac{5}{12} \mathcal{D}_{4}
    \biggr)
    + \CF \TF^2 \nl^2
    \biggl(
        \mathcal{D}_0
        \biggl[
            \frac{29 \pi^2}{324}
            -\frac{581}{648}
        \biggr]
        \nonumber \\ &
        +\mathcal{D}_1
        \biggl[
            \frac{275}{324}
            -\frac{\pi ^2}{27}
        \biggr]
        + \frac{1}{36} \mathcal{D}_2
        -\frac{7}{27} \mathcal{D}_3
    \biggr)
    + \Nc^3
    \biggl(
        -\frac{1}{64} \mathcal{D}_{5}
        -\frac{325}{768} \mathcal{D}_{4}
        +\mathcal{D}_{3}
        \biggl[
            -\frac{8323}{3456}
            -\frac{\pi ^2}{48}
        \biggr]
        \nonumber \\ &
        +\mathcal{D}_2
        \biggl[
            -\frac{\zeta_3}{8}
            -\frac{43 \pi ^2}{192}
            +\frac{6293}{9216}
        \biggr]
        +\mathcal{D}_1
        \biggl[
            \frac{\zeta_3}{48}
            -\frac{101 \pi ^4}{5760}
            +\frac{55 \pi^2}{2304}
            +\frac{221909}{41472}
        \biggr]
        \nonumber \\ &
        +\mathcal{D}_0
        \biggl[
            -\frac{1}{24}\pi ^2 \zeta _3
            +\frac{31 \zeta _3}{192}
            -\frac{\zeta_5}{8}
            -\frac{4673 \pi ^4}{69120}
            +\frac{147241 \pi^2}{82944}
            -\frac{397315}{165888}
        \biggr]
    \biggr)
    \nonumber \\ &
    + \mathcal{O}(\Nc^2,\nh)
    ~,
    \label{eq::spec_plus}
\end{align}
\end{small}%
with $\mathcal{D}_i = \left[ \ln^i(1-x)/(1-x) \right]_{+}$.
The formula for the regular parts are rather lengthy and given in Appendix~\ref{app::formula}.
\begin{figure}[t]
    \begin{center}
    \includegraphics[width=0.8\textwidth]{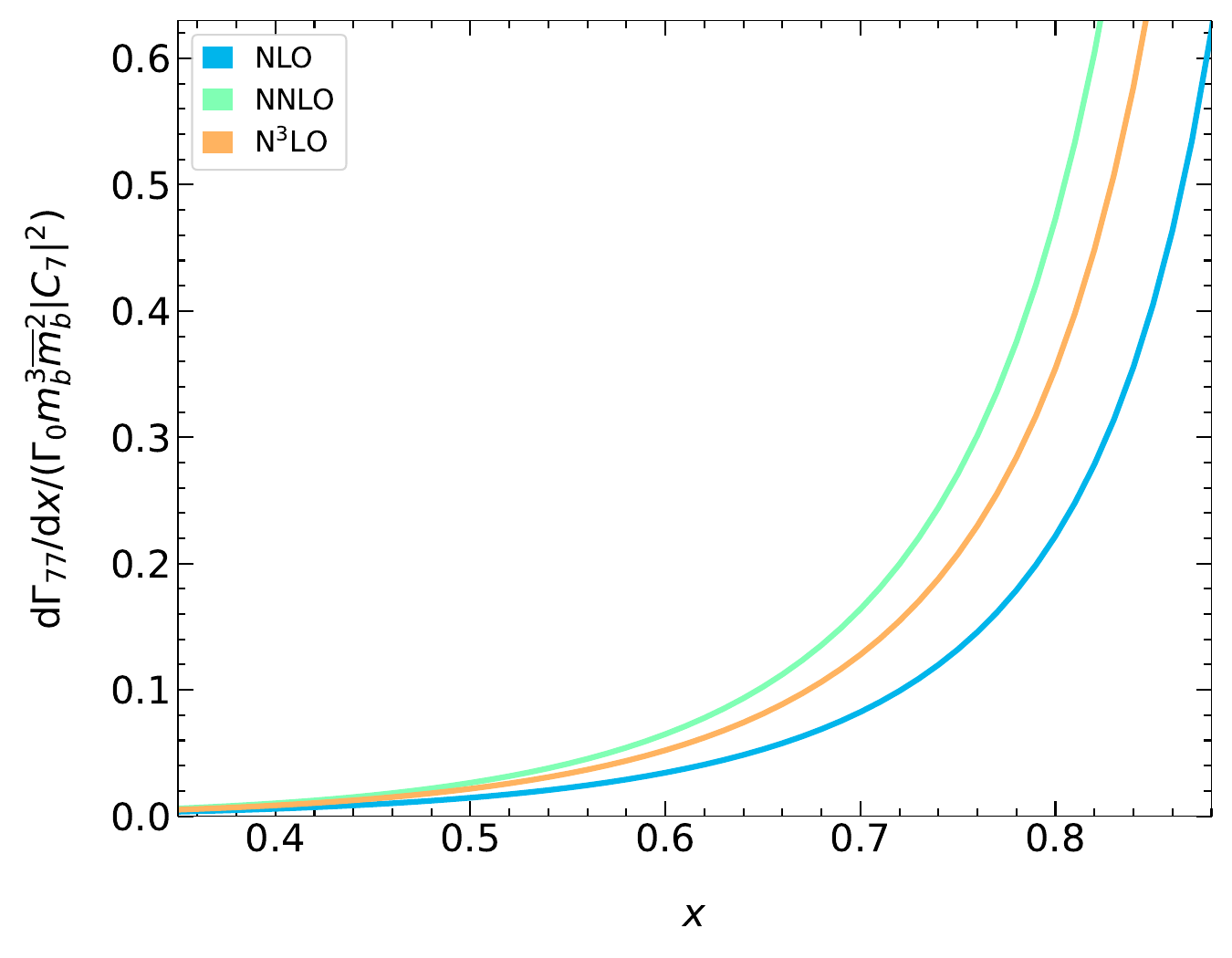}
     \end{center}
    \caption{Normalized spectrum as a function of $x=2E_\gamma/m_b$ in the on-shell scheme. For this plot a fixed value $\alpha_s=0.22$ has been chosen.}  \label{fig::norm_spec}
\end{figure}
In Fig.~\ref{fig::norm_spec} we show the normalized spectrum as a function of $x$ in the pole scheme. This extends Fig.~1 of Ref.~\cite{Melnikov:2005bx} to N$^3$LO.
The plots have been obtained by using the expansions around $x=0$ and $x=1$, which are available in the ancillary files to this paper~\cite{progdata}.
The third-order corrections
amount to approximately 50\% of the NNLO corrections and have
the opposite sign.
We refrain from performing a scheme transformation for the spectrum since
it is to a large extent dominated by the
singularity for $x\to 1$. Instead, we
will discuss in the following subsection the total decay rate, which is
obtained by integrating over the photon
energy.

\subsection{\label{sub::total}Total rate in different renormalization schemes}

We obtain the total decay rate by analytically integrating over the spectrum and expressing the occurring iterated integrals at argument $1$ in terms of transcendental constants,
as described in Section~\ref{sec::details}.

If we choose
the on-shell scheme for the bottom quark mass, the $O_7$ contribution to the total rate for the decay $B \to X_s \gamma$ can be cast in the form
\begin{equation}
    \Gamma_{77} =
    \Gamma_0
    \, m_b^3  \, \overline{m}_b^2(m_b)
    \, |C_7(m_b)|^2
    \Bigg[
    1
    +\frac{\alphas}{\pi} X_1
    +\left(\frac{\alphas}{\pi} \right)^2
    X_2
    +\left(\frac{\alphas}{\pi} \right)^3
    X_3
    \Bigg],
\end{equation}
\noindent with $\Gamma_0 = \alphaem \GF^2  |V_{tb} V_{ts}^\star|^2 / 32\pi^4$.
Since the plus distributions integrate to zero for the decay rate,
the coefficients $X_i$ are obtained by
\begin{align}
    X_{i} &= \int\limits_{0}^{1} \rd x \,Y_{i}
    \,\,=\,\, Y_{i}^{\delta} + \int\limits_{0}^{1} \rd x \,Y_{i}^{\text{reg}}~.
\end{align}
For $\mu=m_b$ we have
\begin{small}
\begin{align}
    X_1 &=
    \CF \left( \frac{4}{3} - \frac{\pi^2}{3}\right),
    \\
    X_2 &=
    \nl \TF \CF \left(
    -\frac{251}{432}
    +\frac{8 \pi ^2}{27}
    +\zeta_3
    \right)
    +\nh \TF \CF
    \left(
    +\frac{3563}{648}
    -\frac{29   \pi ^2}{54}
    -\frac{\zeta_3}{3}
    \right)
    \notag \\ & \quad
    +\CA \CF\left(
    -\frac{1333}{3456}
    +\frac{119 \pi   ^2}{1728}
    -\frac{47 \zeta_{3}}{96}
    -\frac{27}{16} \pi ^2 \log (2)
    +\frac{43 \pi ^4}{1440}
    \right)
    \notag \\ & \quad
    +\CF^2 \left(
    \frac{2825}{288}
    -\frac{319 \pi ^2}{96}
    -\frac{217 \zeta_{3}}{48}
    +\frac{27}{8} \pi ^2 \log (2)
    +\frac{53 \pi^4}{720}
    \right),
    \\
    X_3 &=
    \Nc^3 \left(
    \frac{5678117}{373248}
    -\frac{305723 \pi ^2}{62208}
    -\frac{13177 \zeta_3}{864}
    +\frac{3017   \pi ^4}{10368}
    +\frac{7 \pi ^2 \zeta_3}{72}
    +\frac{925 \zeta_5}{96}
    -\frac{1271 \pi ^6}{181440}
    -\frac{3 \zeta_3^2}{32}\right)
    \notag \\ & \quad
    +\nl^2 \CF \TF^2 \left(
    \frac{5519}{11664}
    -\frac{121 \pi ^2}{243}
    +\frac{14 \zeta_3}{27}
    -\frac{8 \pi   ^4}{405}
    \right)
    - \nh^2 \CF \TF^2 \left(
    \frac{374}{81}
    -\frac{2   \pi ^2}{15}
    -\frac{26 \zeta_3}{9}
    +\frac{\pi ^4}{270}
    \right)
    \notag \\ & \quad
    +\nl \nh \TF^2 \CF
    \biggl(
        -\frac{557}{3888}
        -\frac{505 \pi }{216\sqrt{3}}
        +\frac{469 \pi ^2}{729}
        -\frac{800 \zeta_{3}}{81}
        -\frac{256 \pi ^3}{243\sqrt{3}}
        -\frac{7 \pi ^4}{2430}
        +\frac{128 \pi  \psi^{(1)}\left(\frac{1}{3}\right)}{81 \sqrt{3}}
    \biggr)
    \notag \\ & \quad
    +\nl \TF \CA \CF
    \biggl(
        -\frac{c_{1}}{12}
        -\frac{19 c_{2}}{36}
        -\frac{c_{3}}{3}
        +\frac{2}{15}\pi \text{Im}\left(\text{Li}_3\left(\frac{\ri}{\sqrt{3}}\right)\right)
        -\frac{70 \text{Li}_4\left(\frac{1}{2}\right)}{9}
        -\frac{10187\zeta_{5}}{2592}
        \notag \\ & \quad
        -\frac{487 \pi ^2 \zeta_{3}}{2592}
        +\frac{5\pi  \zeta_{3}}{24 \sqrt{3}}
        +\frac{433 \zeta_{3}}{192}
        +\frac{5 \pi ^5}{729 \sqrt{3}}
        +\frac{1495 \pi^4}{11664}
        +\frac{655 \pi ^3}{648 \sqrt{3}}
        +\frac{28081 \pi^2}{15552}
        +\frac{661 \pi }{576\sqrt{3}}
        \notag \\ & \quad
        -\frac{1269163}{373248}
        -\frac{35 \log^4(2)}{108}
        -\frac{1}{360} \pi ^2 \log ^2(3)
        -\frac{395 \pi ^2\log ^2(2)}{1296}
        -\frac{19 \pi  \log ^2(2)}{72\sqrt{3}}
        +\frac{19 \log ^2(2)}{192}
        \notag \\ & \quad
        -\frac{\pi ^3 \log (3)}{54\sqrt{3}}
        +\frac{\pi  \log (3)}{24 \sqrt{3}}
        +\frac{253}{72}\pi ^2 \log (2)
        -\frac{\pi  \psi^{(3)}\left(\frac{2}{3}\right)}{324 \sqrt{3}}
        -\frac{\psi^{(1)}\left(\frac{1}{3}\right)^2}{108}
        +\frac{\pi ^3 \psi^{(1)}\left(\frac{1}{3}\right)}{486 \sqrt{3}}
        \notag \\ & \quad
        +\frac{1}{81}\pi ^2 \psi ^{(1)}\left(\frac{1}{3}\right)
        -\frac{629 \pi \psi ^{(1)}\left(\frac{1}{3}\right)}{432 \sqrt{3}}
        -\frac{\psi^{(1)}\left(\frac{1}{3}\right)}{36}
        +\frac{\pi  \log (3) \psi^{(1)}\left(\frac{1}{3}\right)}{36 \sqrt{3}}
    \biggr)
    \notag \\ & \quad
    + \nl \TF \CF^2
    \biggl(
        \frac{c_{1}}{6}
        +\frac{19 c_{2}}{18}
        +\frac{2 c_{3}}{3}
        -\frac{4}{15}\pi \text{Im}\left(\text{Li}_3\left(\frac{\ri}{\sqrt{3}}\right)\right)
        +\frac{140 \text{Li}_4\left(\frac{1}{2}\right)}{9}
        -\frac{613\zeta_{5}}{1296}
        \notag \\ & \quad
        +\frac{127 \pi ^2 \zeta_{3}}{1296}
        -\frac{5\pi  \zeta_{3}}{12 \sqrt{3}}
        +\frac{7795 \zeta_{3}}{864}
        -\frac{10 \pi ^5}{729 \sqrt{3}}
        -\frac{3599 \pi^4}{7290}
        -\frac{655 \pi ^3}{324 \sqrt{3}}
        +\frac{29 \pi^2}{4}
        -\frac{661 \pi }{288\sqrt{3}}
        \notag \\ & \quad
        -\frac{274397}{20736}
        +\frac{35 \log^4(2)}{54}
        +\frac{1}{180} \pi ^2 \log ^2(3)
        +\frac{395}{648}\pi ^2 \log ^2(2)
        +\frac{19 \pi  \log ^2(2)}{36\sqrt{3}}
        -\frac{19 \log ^2(2)}{96}
        \notag \\ & \quad
        +\frac{\pi ^3 \log (3)}{27\sqrt{3}}
        -\frac{\pi  \log (3)}{12 \sqrt{3}}
        -\frac{253}{36}\pi ^2 \log (2)
        +\frac{\pi  \psi^{(3)}\left(\frac{2}{3}\right)}{162 \sqrt{3}}
        +\frac{\psi^{(1)}\left(\frac{1}{3}\right)^2}{54}
        -\frac{\pi ^3 \psi^{(1)}\left(\frac{1}{3}\right)}{243 \sqrt{3}}
        \notag \\ & \quad
        -\frac{2}{81}\pi ^2 \psi ^{(1)}\left(\frac{1}{3}\right)
        +\frac{629 \pi \psi ^{(1)}\left(\frac{1}{3}\right)}{216 \sqrt{3}}
        +\frac{\psi^{(1)}\left(\frac{1}{3}\right)}{18}
        -\frac{\pi  \log (3) \psi^{(1)}\left(\frac{1}{3}\right)}{18 \sqrt{3}}
    \biggr)
    +O(\Nc^2,\nh)~.
\end{align}
\end{small}

\noindent The results for $X_1$ and $X_2$ are in agreement with Ref.~\cite{Blokland:2005uk};
the result for $X_3$ is new.
Computer-readable expressions
for general renormalization scale $\mu$ can
be found in the ancillary files~\cite{progdata}.

The numerical evaluation for $\mu=m_b$ leads to
\begin{align}
\Gamma_{77} &=
    \Gamma_0
    m_b^3 \overline{m}_b^2(m_b)
    |C_7(m_b)|^2
    \Bigg[
    1
    - 2.60871 \left( \frac{\alphas}{\pi} \right)
    \nonumber \\ &
    + \left(\frac{\alphas}{\pi} \right)^2
    \biggl(
        -32.4531
        - 0.135048 \nh + 2.36358 \nl
    \biggr)
    \nonumber \\ &
    + \left(\frac{\alphas}{\pi} \right)^3
    \biggl(
        -513.126 - 0.0631674 \nh^2  + 69.174 \nl
        - 0.0365688 \nh \nl - 1.91406 \nl^2
        \nonumber \\ &
        +\mathcal{O}(\Nc^2,\nh)
    \biggr)
    \Bigg]
    \nonumber\\
    &\overset{\nl=4,\nh=1}{=}
    \Gamma_0( m_b^3 \overline{m}_b^2(m_b) ) |C_7(m_b)|^2
    \Bigg[1 - 2.61 \frac{\alphas}{\pi}
    - 23.13 \left(\frac{\alphas}{\pi} \right)^2
    - 267.26 \left(\frac{\alphas}{\pi} \right)^3\Bigg],
    \label{eq::Gam_pole}
\end{align}
which manifests the usual poor convergence of the perturbative
series in the on-shell scheme.

One caveat of our calculation lies in the fact that we have only computed the large-$\Nc$, light-fermion, and $\nh^2$ contributions.
One can therefore ask if the subleading-$\Nc$ terms could partially cancel the large N${}^{3}$LO contributions observed here.
At NNLO this does not seem to be the case.
Naively the subleading-$\Nc$ terms come with a suppression factor $1/\Nc$ and therefore contribute a $\sim 30$\% shift.
In practice we rather observe a correction of about $10$\%.
We therefore do not expect the subleading-$\Nc$ terms to substantially alter the observations in this and the following Sections.

In the following, we investigate the convergence and scale dependence of
the total rate in the kinetic~\cite{Bigi:1996si} and the MSR schemes~\cite{Hoang:2008yj,Hoang:2017suc}.

\subsubsection*{Kinetic scheme}
We transform the pole mass $m_b\equiv m_b^{\rm pole}$ to the kinetic scheme
and we take into account the power-suppressed leading-order contributions to $\Gamma_{77}$
stemming from $\mu_\pi^2$ and $\rho_D^3$ (see for instance Ref.~\cite{Bauer:1997fe}).
Their transformation to the kinetic scheme induces perturbative contributions which we have to take into account.
We use the conversion formula between $m_b^{\rm pole}$ and $m_b^{\rm kin}$
and the expressions for $\mu_\pi^2\vert_\mathrm{pert}$ and $\rho_D^3|_\mathrm{pert}$ up to third order from Refs.~\cite{Czarnecki:1997sz,Fael:2020iea,Fael:2020njb}.
We use the formulae from Ref.~\cite{Fael:2020njb}, which correspond to massless
charm quarks, as this is the approximation which we also use in the calculation of the photon spectrum.
Therefore, to convert the on-shell mass to the kinetic mass
we consider the expressions in Appendix A of Ref.~\cite{Fael:2020njb}
with $\nl=4$ and $\alphas^{(4)}$ and apply the decoupling
relation to express the
mass relation in terms of $\alphas^{(5)}$. For the decoupling scale we use  $\overline{m}_b(\mu)$.
Converting only the on-shell mass to the kinetic scheme
and reexpanding up to third order, we obtain
with $m_b^{\rm kin} (1 \, \mathrm{GeV})=4.55$~GeV
and $\overline{m}_b (m_b^{\rm kin} ) = 4.14$~GeV
\begin{align}
\Gamma_{77} &=
    \Gamma_0\,
     |C_7(m_b^{\rm kin})|^2 \,
    (m_b^{\rm kin})^3 \,  \overline{m}_b^2(m_b^{\rm kin} ) \,
    \Bigg[1 - 1.29 \frac{\alphas}{\pi}
    - 8.75 \left(\frac{\alphas}{\pi} \right)^2
    - 80.82 \left(\frac{\alphas}{\pi} \right)^3\Bigg] \nonumber\\
    &=
    \Gamma_0 \, |C_7(m_b^{\rm kin})|^2 \,
 (m_b^{\rm kin})^3 \,  \overline{m}_b^2(m_b^{\rm kin} ) \,
    \Big[
    1 - 0.090_{\alphas} - 0.043_{\alphas^2} - 0.028_{\alphas^3}
    \Big],
    \label{eq::Gam_kinMS}
\end{align}
where $\alphas \equiv \alphas^{(5)}(m_b^{\rm kin})$.
We observe a reasonable behaviour of the perturbative series,
although with NNLO and N$^3$LO effects still in the range of a few percent,
beyond the sub-percent effect that one would naively expect.

In a next step we transform all instances of $\overline{m}_b^2$
to the kinetic scheme using the expressions from Ref.~\cite{Fael:2020njb}
and obtain
\begin{align}
\Gamma_{77} &=
    \Gamma_0 |C_7(m_b^{\rm kin})|^2 \,
    (m_b^{\rm kin})^5 \,
    \Bigg[1 - 3.11 \frac{\alphas}{\pi}
    - 17.13 \left(\frac{\alphas}{\pi} \right)^2
    - 122.22 \left(\frac{\alphas}{\pi} \right)^3\Bigg].
\end{align}
The comparison to Eq.~(\ref{eq::Gam_kinMS}) shows that
the pertubative expansion is somewhat worse, although  it is still better than in the pole scheme, see Eq.~(\ref{eq::Gam_pole}).

To study the dependence of the N$^3$LO total rate on the renormalization
scale $\mu$,\footnote{We evaluate $\alphas(\mu)$, $\overline{m}(\mu)$, and $C_7(\mu)$ at the same scale.} one would in principle require the evolution
of the Wilson coefficient $C_7(\mu)$ at N$^3$LL accuracy.
However, the mixing between $O_7$ and $O_8$ is
currently known only up to NNLL~\cite{Misiak:1994zw,Gorbahn:2005sa},
as is the mixing of current-current and penguin operators
into $O_7$ and $O_8$~\cite{Czakon:2006ss}.
Moreover, the matching conditions are available only up to
$O(\alphas^2)$~\cite{Misiak:2004ew}.
As a consequence, any estimate of the renormalization-scale uncertainty
would not be limited by our N$^3$LO evaluation of the matrix element,
but by the incomplete knowledge of the running and matching of the operator
$O_7$.

We therefore adopt a simplified approach.
We evaluate the Wilson coefficient $C_7$ at the scale $\mu = m_b$
and study the scale dependence of the rate by assuming that
the evolution of
\begin{equation}
    C_{\mathrm{T}}(\mu) = C_7(\mu) \overline{m}_b(\mu)
\end{equation}
is governed solely by the anomalous dimension of the tensor current,
known up to $O(\alphas^4)$ from Refs.~\cite{Gracey:2000am,Baikov:2006ai,Gracey:2022vqr}.\footnote{Eq.~(3.2) of Ref.~\cite{Gracey:2022vqr} contains a typo: the colour factor $d_A^{abcd} d_A^{abcd}$ at order $\alphas^4$ must be replaced by $N_f d_F^{abcd} d_F^{abcd}$
to have agreement with Ref.~\cite{Baikov:2006ai}.
}
The $\mu$ dependence of $C_{\mathrm{T}}$ is given by the solution of
the renormalization group equation
\begin{equation}
   \mu \frac{\rd C_{\mathrm{T}}(\mu)}{\rd \mu} = \gamma_{\mathrm{T}}(\alphas) C_{\mathrm{T}}(\mu),
\end{equation}
which reads
\begin{equation}
    C_{\mathrm{T}}(\mu) = C_{\mathrm{T}}(\mu_0) \exp
    \left[
    \int_{\alphas(\mu_0)}^{\alphas(\mu)}
    \frac{\gamma_{\mathrm{T}}(\alphas)}{2 \beta(\alphas)}
    \mathrm{d}\alphas
    \right]
    \label{eqn:RGECT}
\end{equation}
with
the QCD beta function and the anomalous dimension of the tensor current given by
\begin{align}
    \beta(\alphas) &= - \sum_{n=0} \beta_n \left(\frac{\alphas}{\pi}\right)^{n+2}, &
    \gamma_{\mathrm{T}}(\alphas) &= \sum_{n=0} \gamma_n \left(\frac{\alphas}{\pi}\right)^{n+1}\,.
\end{align}
Inserting the expansions for $\gamma_{\mathrm{T}}(\alphas)$ and
$\beta(\alphas)$ into Eq.~\eqref{eqn:RGECT} and expanding in $\alphas$ gives:
\begin{align}
   C_{\mathrm{T}}(\mu) &= C_{\mathrm{T}}(\mu_0)
   \left[
   1+ J_1 \frac{\alphas(\mu)}{\pi}
   + J_2 \left( \frac{\alphas(\mu)}{\pi} \right)^2
   + J_3 \left( \frac{\alphas(\mu)}{\pi} \right)^3
   \right]
   \left( \frac{\alphas(\mu_0)}{\alphas(\mu)} \right)^{\frac{\gamma_0}{2 \beta_0}} \notag \\
   & \times
   \left[
   1 - J_1 \frac{\alphas(\mu_0)}{\pi}
   - (J_2-J_1^2) \left( \frac{\alphas(\mu_0)}{\pi} \right)^2
   -(J_3+J_1^3-2 J_1 J_2) \left( \frac{\alphas(\mu_0)}{\pi} \right)^3
   \right]\,,
   \label{eqn:runningCT}
\end{align}
where
\begin{align}
    J_1 &= \frac{\beta_1 \gamma_0}{2 \beta_0^2}-\frac{\gamma_1}{2 \beta_0} \,, \notag \\
    J_2 &= \frac{\beta _1^2 \gamma _0^2}{8 \beta _0^4}-\frac{\beta _1^2
   \gamma _0}{4 \beta _0^3}-\frac{\beta _1 \gamma _0 \gamma _1}{4
   \beta _0^3}+\frac{\beta _1 \gamma _1}{4 \beta
   _0^2}+\frac{\gamma _1^2}{8 \beta _0^2}+\frac{\beta _2 \gamma
   _0}{4 \beta _0^2}-\frac{\gamma _2}{4 \beta _0} \, , \notag \\
   J_3 &=
   \frac{\beta _1^3 \gamma _0^3}{48 \beta _0^6}-\frac{\beta _1^3 \gamma _0^2}{8 \beta _0^5}+\frac{\beta _1^3 \gamma _0}{6 \beta _0^4}-\frac{\beta
   _1^2 \gamma _0^2 \gamma _1}{16 \beta _0^5}+\frac{\beta _1^2 \gamma _0 \gamma _1}{4 \beta _0^4}-\frac{\beta _1^2 \gamma _1}{6 \beta
   _0^3}+\frac{\beta _2 \beta _1 \gamma _0^2}{8 \beta _0^4}+\frac{\beta _1 \gamma _0 \gamma _1^2}{16 \beta _0^4}-\frac{\beta _1 \gamma _1^2}{8
   \beta _0^3}
   \notag \\ & \quad
    -\frac{\beta _2 \beta _1 \gamma _0}{3 \beta _0^3}-\frac{\beta _1 \gamma _0 \gamma _2}{8 \beta _0^3}+\frac{\beta _1 \gamma _2}{6
   \beta _0^2}-\frac{\gamma _1^3}{48 \beta _0^3}+\frac{\beta _3 \gamma _0}{6 \beta _0^2}+\frac{\beta _2 \gamma _1}{6 \beta _0^2}-\frac{\beta _2
   \gamma _0 \gamma _1}{8 \beta _0^3}+\frac{\gamma _1 \gamma _2}{8 \beta _0^2}-\frac{\gamma _3}{6 \beta _0} \, .
\end{align}
\begin{figure}[t]
    \centering
    \begin{tabular}{c}
    \includegraphics[width=0.8\linewidth]{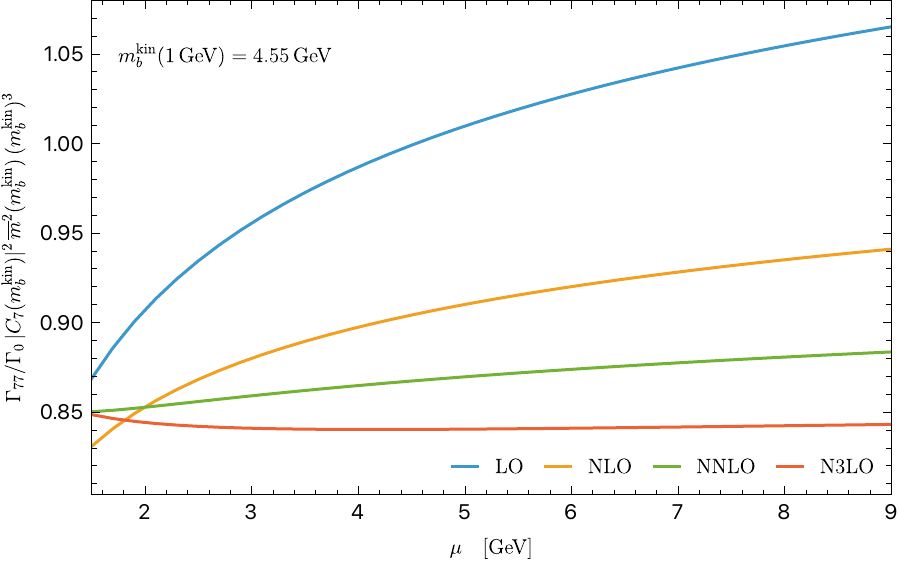}
    \end{tabular}
    \caption{The total partonic decay rate in the kinetic mass scheme
     as a function of the renormalization scale $\mu$.
     }
    \label{fig:mudependenceKIN3MS2}
\end{figure}
In Fig.~\ref{fig:mudependenceKIN3MS2} we present the total decay rate in the
kinetic scheme as a function of the renormalization scale $\mu$.
Specifically, we consider the rate $\Gamma_{77}$ normalized to
$C_{\mathrm{T}}^2 = |C_7|^2 \, \overline{m}_b^2$, evaluated at the reference scale
$\mu = m_b^{\mathrm{kin}}$,
and to an additional mass factor $(m_b^{\mathrm{kin}})^3$.
With this normalization, the dependence of the rate on $\mu$ can be studied
without fixing the numerical value of $C_7(m_b^{\mathrm{kin}})$.

The N$^n$LO curves are obtained by multiplying our result for the matrix element up to $\alphas^n$, evaluated at the scale $\mu$, by $C_{\mathrm{T}}(\mu)$
as given in Eq.~\eqref{eqn:runningCT} considering the RGE
resummation up to N$^n$LL, without reexpanding the result in $\alphas$.
For the numerical evaluation of $\alphas^{(5)}(\mu)$ and
$\overline{m}_b(\mu)$ we employ \texttt{RunDec}~\cite{Chetyrkin:2000yt,Herren:2017osy},
using $\alphas^{(5)}(M_Z) = 0.118$ and
$\overline{m}_b^{(5)}(\overline{m}_b) = 4.2\,\mathrm{GeV}$ as input values.

Over the entire range $m_b^{\mathrm{kin}}/2 \le \mu < 2 m_b^{\mathrm{kin}}$,
the sensitivity of the rate to the renormalization scale is progressively
reduced with the inclusion of higher-order corrections.
The $7\%$ variation observed at leading order decreases to $4\%$, $1.5\%$,
and $0.2\%$ upon including NLO, NNLO, and N$^3$LO corrections in the matrix
element and in the running, respectively.
On the other hand, we find only a marginal overlap of the uncertainty bands
obtained from a simple variation of $\mu$.
In particular, the N$^3$LO prediction does not overlap with the NNLO error band
for $m_b^{\mathrm{kin}}/2 \le \mu < 2 m_b^{\mathrm{kin}}$.
As a consequence, the NNLO uncertainty band obtained from scale variation does not fully account for the size of the genuine higher-order contributions,
indicating that, at NNLO, the $\mu$ variation underestimates the impact of
missing perturbative corrections in this observable.

\subsubsection*{MSR scheme}
As an alternative to the kinetic mass scheme
we transform the pole mass
into the MSR scheme. We follow Ref.~\cite{Hoang:2017suc}
and use the so-called ``practical MSR'' mass, which is
introduced in Section~2.3 of that reference.

With $\mu_s=m_b^{\rm MSR}$, $m_b^{\rm MSR} (1\, \mathrm{GeV})=4.66$~GeV,
and $\overline{m}_b (m_b^{\rm MSR})=4.122$~GeV we obtain
\begin{align}
\Gamma_{77} &=
    \Gamma_0
    |C_7(m_b^{\rm MSR})|^2
    \, (m_b^{\rm MSR})^3  \, \overline{m}_b^2(\mu_s)  \,
    \Bigg[
    1. - 1.75 \frac{\alphas}{\pi}
    - 13.30 \left(\frac{\alphas}{\pi} \right)^2
    - 121.50 \left(\frac{\alphas}{\pi} \right)^3
    \Bigg]
    \notag \\ &
    =
    \Gamma_0
    |C_7(m_b^{\rm MSR})|^2
    \, (m_b^{\rm MSR})^3  \, \overline{m}_b^2(\mu_s)  \,
    \Big[
    1 - 0.121_{\alphas} - 0.064_{\alphas^2} - 0.040_{\alphas^3}
    \Big].
\end{align}
Also for this MSR scheme we observe a reasonable behaviour
of the perturbative series,
although with larger
corrections as in the kinetic mass scheme, see Eq.~(\ref{eq::Gam_kinMS}).
\begin{figure}[t]
    \centering
    \begin{tabular}{c}
    \includegraphics[width=.8\linewidth]{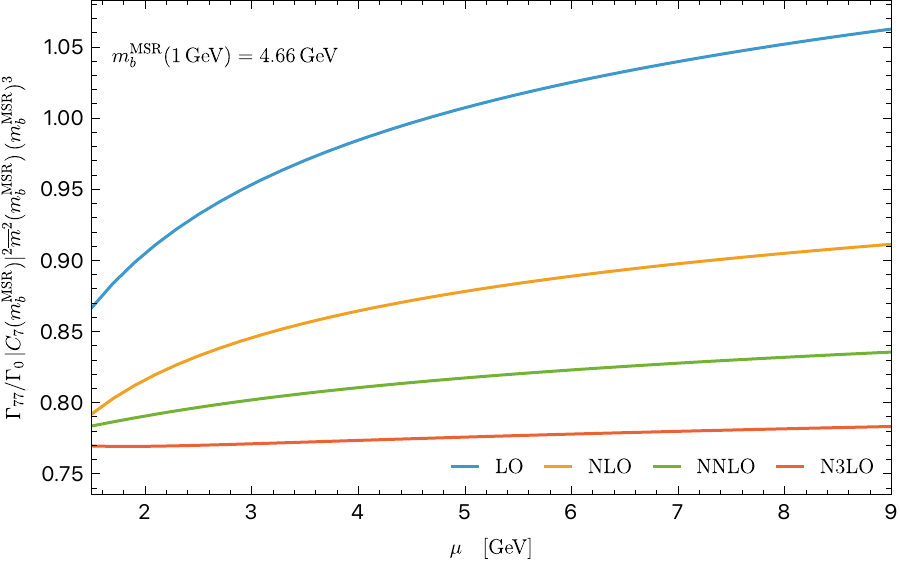}
    \end{tabular}
    \caption{The total partonic decay rate with $m_b$ in the MSR scheme
     as a function of the renormalization scale $\mu$.
     }
    \label{fig:mudependenceMSR3MS2}
\end{figure}
Following the same approach as the kinetic scheme, we show in Fig.~\ref{fig:mudependenceMSR3MS2} the dependence of the rate
on $\mu$ in the range $m_b^\mathrm{MSR} / 2 \le \mu \le 2 m_b^\mathrm{MSR} $.
Also in this case we observe a reduction of the 7\% scale uncertainty
of the LO. The inclusion of NLO, NNLO, and N$^3$LO corrections
reduces the uncertainty based on scale variation to 4\%, 2\%, and 0.7\%.
However, again the size of the higher-order corrections is not captured by this simple estimate.

\subsection{Branching ratio with photon energy cut}

\begin{table}[t]
    \centering
    \begin{tabular}{c|c|c|c|c}
    \hline \hline
   $E_\mathrm{cut}$  &
   $X_1^{\rm cut}$ &
   $X_2^{\rm cut}$ &
   $X_3^{\rm cut}$ &
   $\Gamma_{77}^{\rm norm}$
   \\
   \hline
   $0.5 \, \text{GeV}$  & $-1.2909$ & $-8.75$ & $-80.84$ & $0.840$ \\
   $1.0 \, \text{GeV}$  & $-1.3016$ & $-8.81$ & $-81.27$   & $0.839$ \\
   $1.6 \, \text{GeV}$  & $-1.4264$ & $-9.56$ & $-87.42$  & $0.824$ \\
   $1.8 \, \text{GeV}$  & $-1.5939$ & $-10.55$ & $-9.613$  & $0.805$ \\
   $2.0 \, \text{GeV}$  & $-2.0434$ & $-12.84$ & $-117.14$  & $0.755$ \\
   $2.2 \, \text{GeV}$  & $-4.2763$ & $-11.28$ & $-155.90$ & $0.594$
   \\ \hline
    \end{tabular}
    \caption{Coefficients of the perturbative expansion of $\Gamma_{77}(E_\mathrm{cut})$
    in the kinetic scheme, keeping the factor $\overline{m}_b^2$.
    The LO contribution is independent on $E_\mathrm{cut}$ and
    always equal to one.
    $\Gamma_{77}^{\rm norm}$ is defined such that the factor
    $\Gamma_0
    |C_7(m_b^{\rm kin})|^2
    \, (m_b^{\rm kin})^3  \, \overline{m}_b^2(m_b^{\rm kin})$
    is divided out. For the renormalization scale we have chosen $\mu=m_b^{\rm kin}$.}
    \label{tab:BRcutKIN}
\end{table}

\begin{table}[t]
    \centering
    \begin{tabular}{c|c|c|c|c}
    \hline \hline
   $E_\mathrm{cut}$  &
   $X_1^{\rm cut}$ &
   $X_2^{\rm cut}$ &
   $X_3^{\rm cut}$ &
   $\Gamma_{77}^{\rm norm}$
   \\
   \hline
   $0.5 \, \text{GeV}$  & $-1.7508$ & $-13.31$ & $-121.52$ & $0.775$ \\
   $1.0 \, \text{GeV}$  & $-1.7603$ & $-13.36$ & $-121.95$ & $0.774$ \\
   $1.6 \, \text{GeV}$  & $-1.8676$ & $-14.06$ & $-128.04$ & $0.761$ \\
   $1.8 \, \text{GeV}$  & $-2.0050$ & $-14.97$ & $-136.86$ & $0.744$ \\
   $2.0 \, \text{GeV}$  & $-2.3468$ & $-17.19$ & $-159.69$ & $0.702$\\
   $2.2 \, \text{GeV}$  & $-3.6149$ & $-23.31$ & $-225.13$ & $0.563$
   \\ \hline
    \end{tabular}
    \caption{Coefficients of the perturbative expansion of $\Gamma(E_\mathrm{cut})$
    in the MSR scheme.
    The LO contribution is independent on $E_\mathrm{cut}$ and
    always equal to one.
    $\Gamma_{77}^{\rm norm}$ is defined such that the factor
    $\Gamma_0
    |C_7(m_b^{\rm MSR})|^2
    \, (m_b^{\rm MSR})^3  \, \overline{m}_b^2(m_b^{\rm MSR})$
    is divided out. For the renormalization scale we have chosen $\mu=m_b^{\rm MSR}$.
    }
    \label{tab:BRcutMSR}
\end{table}

When integrating over the spectrum, it is possible to introduce a lower cut on the photon energy and define the cut-dependent total decay rate as
\begin{equation}
    \Gamma(E_\mathrm{cut}) =
    \int_{E_\mathrm{cut}}^{m_b/2} \frac{\rd \Gamma}{\rd E} \, \rd E.
    \label{eq::Gamma_cut}
\end{equation}
For $E_{\rm cut}=0$ we obtain the results presented in Section~\ref{sub::total}.
If $E_{\rm cut}>0$, it is
convenient to split the integrand in Eq.~(\ref{eq::Gamma_cut}) into the singular and regular part.
The former gets contributions from the $\delta$ functions which is identical for all
values of $E_{\rm cut}$.
This part can be taken over from Section~\ref{sub::total}.
For the contribution of the plus distributions to the singular part
one has to take care of the fact that for the lower bound of the
integral in Eq.~(\ref{eq::Gamma_cut}) we have $E_{\rm cut}>0$.  This
leads to an additional term which can be expressed as an integral over the region
$E\in[0,E_{\rm cut}]$ where the plus distributions are regular, see Eq.~(\ref{eq:plus-distr-ycut}).  This
contribution contains an explicit dependence on the bottom quark mass,
even from the integration boundary, which has to be taken into account
when performing scheme changes. This is also the case for the
regular part.

\begin{figure}[t]
    \centering
    \begin{tabular}{c}
    \includegraphics[width=0.79\linewidth]{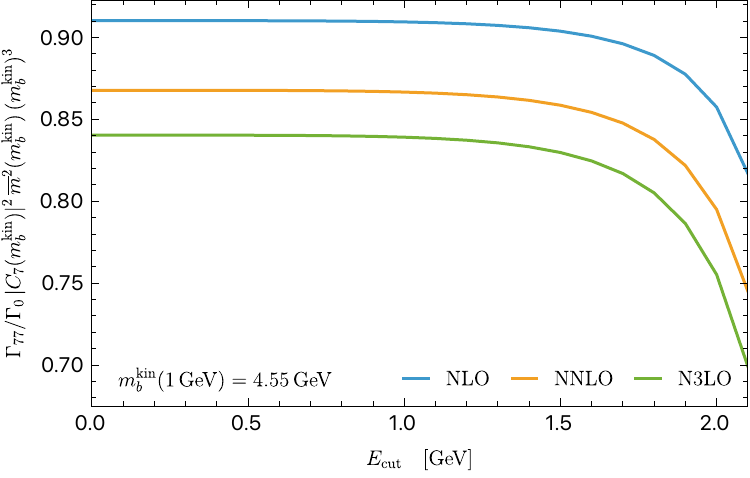} \\
    \includegraphics[width=0.79\linewidth]{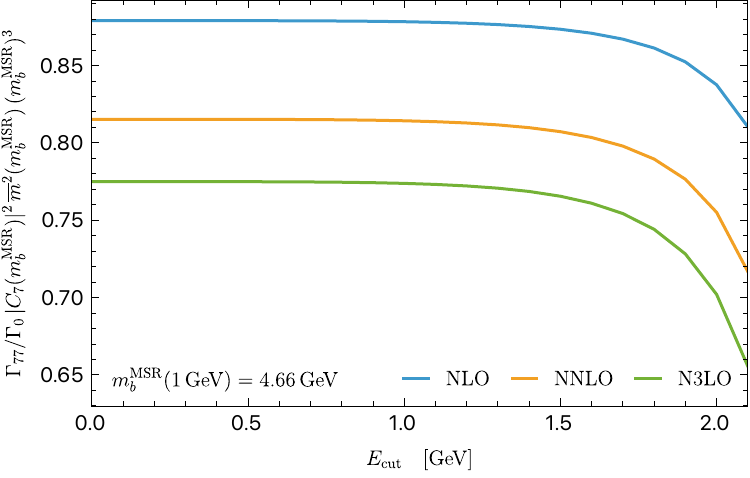}
    \end{tabular}
    \caption{The $O_7$ contribution to the rate $\Gamma$ as a function of
    on the lower cut on the photon energy $E_\mathrm{cut}$ in the kinetic scheme (top)
    and the MSR scheme (bottom).
    }
    \label{fig:BRcutKIN_MSR}
\end{figure}

It is convenient to define
\begin{align}
    X_{i}^{\rm cut} &=
    \int\limits_{x_{\rm cut}}^{1} \rd x \, Y_{i}
\end{align}
with $x_{\rm cut}=2 E_{\rm cut}/m_b$
such that the $E_{\rm cut}$-dependent total decay rate is given by
\begin{align}
    \Gamma_{77}(E_{\rm cut})
    &=\Gamma_0
    |C_7(m_b)|^2
    \, m_b^3  \, \overline{m}_b^2(m_b)
    \left[
        1 + \sum\limits_{i=1}^{3} \left(\frac{\alpha_s}{\pi}\right)^{i} X_{i}^{\rm cut}
    \right]\,.
\end{align}
Our results are shown in Tabs.~\ref{tab:BRcutKIN}
and~\ref{tab:BRcutMSR}
for the kinetic (keeping the factor $\overline{m}_b^2$) and MSR schemes, respectively, for various values
of $E_{\rm cut}$. We observe a similar pattern as in Section~\ref{sub::total}: NLO, NNLO, and N$^3$LO corrections have the same sign and are negative. 
Figure~\ref{fig:BRcutKIN_MSR} 
shows the properly normalized decay rate
$\Gamma_{77}$ as a function of $E_{\rm cut}$.

\subsection{Comparison to previous estimates of the N$^3$LO corrections}

In Ref.~\cite{Dehnadi:2022prz} the photon energy spectrum has been analysed by matching resummed singular contributions in the  peak region, i.e.\ for large values of the photon energy, to fixed order results in the tail region.
In this analysis a couple of fixed-order three-loop contributions were not known explicitly, but treated as nuisance parameters in the fit to data. In particular,
the following two ingredients were
unavailable:
\begin{itemize}
    \item The finite part of the three-loop heavy-to-light form factor, which has been computed in Ref.~\cite{Fael:2024vko}.
    \item The finite part of the three-loop $b \to s \gamma$ spectrum $Y_{3}(x)$,
    which is computed in this paper in the large-$\Nc$ limit, including all light-fermion and $\nh^2$ contributions.
\end{itemize}

In Ref.~\cite{Fael:2024vko}
it was shown that the three-loop contribution to the heavy-to-light form factor is significantly larger than the estimate used in the Ref.~\cite{Dehnadi:2022prz}:
\begin{align}
    h_{3}^{\text{Ref.\,\cite{Dehnadi:2022prz}}} &= 0 \pm 80, & h_{3}^{\text{exact}}&=
    -181.1617381... ~.
\end{align}

With the calculation presented in this work we are in a position to address also the second unknown ingredient.
In Ref.~\cite{Dehnadi:2022prz} the contribution to $Y_3(x)$ is split into a singular and a regular piece,
\begin{align*}
    Y_3(x) &= Y_3^{\text{sing}}(x)
    + Y_3^{\text{reg}}(x)
    ~.
\end{align*}
The singular piece consists of all distributions and can be calculated within SCET from known ingredients.
In Ref.~\cite{Dehnadi:2022prz} the expression is given for SU$(3)$ and we agree with the one- and two-loop expressions.
At three loops we can only compare the terms proportional to powers of $\nl$ and we find agreement as well.

\begin{figure}[t]
    \centering
    \includegraphics[width=0.9\linewidth]{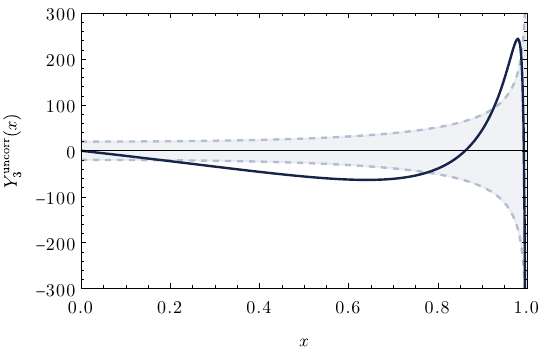}
    \caption{Comparison between the modelling of nuisance parameters related to the, at that point unknown, $Y_{3}^{\text{uncorr}}(x)$ following Ref.~\cite{Dehnadi:2022prz}
    (gray band)
    and the leading-color, light-fermion, and $\nh^2$ results obtained in this paper (black line).
    }
    \label{fig:compare}
\end{figure}

In Ref.~\cite{Dehnadi:2022prz} the prediction for the regular contribution $Y_3^{\text{reg}}(x)$ was done for an effective Wilson coefficient, including contributions from the Wilson coefficients of other operators besides $O_7$ considered in the present paper.
Hence, a direct comparison is not possible.
Instead, we follow the same prescription to derive an estimate for our expression:
The regular piece is split into
\begin{align}
    Y_3^{\text{reg}}(x) &=
    Y_3^{\text{corr}} +
    Y_3^{\text{uncorr}}(x) ~,
\end{align}
with $Y_3^{\text{corr}}=-Y_3^{\text{sing}}(0)$.
This ensures that $Y_3^{\text{uncorr}}(0)=0$~\cite{Dehnadi:2022prz}.
$Y_3^{\text{uncorr}}(x)$ is then modelled via powers of the analytically known spectrum at NLO, $Y_{1}(x)$, via
\begin{align}
    Y_3^{\text{uncorr}}(x) = \sum\limits_{i=0}^{5} c_i \left(\frac{Y_1(x)-Y_1(0)}{4} \right)^i~,
\end{align}
with max($|c_{i}|$) taken to be $\{
20,100,80,10,5,1
\}$.
This ensures that the powers of logarithms in the limit $x \to 1$ are reproduced correctly.
The uncertainty around the central value $0$ is then obtained by varying the six parameters independently and adding the corresponding uncertainties in quadrature.
In Fig.~\ref{fig:compare} we show the region covered by this ansatz as well as our leading-color, light-fermion, and $\nh^2$ result for $Y_3^{\text{uncorr}}(x)$.
We notice that the ansatz provides a decent estimate of the corrections, albeit slightly underestimating the uncertainty.
However, especially in the limit $x \to 1$ the uncertainties become large.
With the results obtained in the present paper, we are able to shrink them significantly to the missing sub-leading color contributions only.

\section{\label{sec::concl}Conclusions}

In this work we consider the photon energy spectrum to the partonic radiative
decay $b\to s \gamma$, induced by the electric dipole operator $O_7$.  We
reproduce the NNLO results available in the literature and compute all
light-fermion and $\nh^2$ contributions as well as the leading term in the large-$\Nc$ expansion
at N$^3$LO.

In our approach, we compute the imaginary part of the forward scattering
amplitude $b\gamma \to b\gamma$, which must be considered up to three
loops. Since the spectrum is divergent at the end point, i.e., for
$E_\gamma \to m_b/2$, it is important to distinguish the individual branches
during the computation, which we realize at the level of the master
integrals. This allows us to isolate the singular behaviour and write the
analytic expressions in terms of $\delta$ and plus distributions and functions
which are regular for $E_\gamma \to m_b/2$.  We provide explicit analytic
results for the spectrum which can be downloaded in computer-readable
from the ancillary files~\cite{progdata}.

Analytic results for the total inclusive decay rate $\Gamma(b\to s\gamma)$ are obtained after
integrating over all kinematically allowed values of $E_\gamma$.  We discuss
the convergence properties in three different renormalization schemes and
observe a significant improvement after applying  a short-distance scheme.  We
also compute the total decay rate as a function of the photon energy cut.

The techniques presented in this work can also be applied to the contributions
from subleading colour factors. In that case, however, also non-planar Feynman
diagrams contribute. Furthermore, the number of master integrals is
significantly higher, which complicates analytical calculations and likely
necessitates numerical methods.  Our method can also be used to calculate the
contribution from the chromomagnetic dipole operator $O_8$ to the photon
energy spectrum.

\section*{Acknowledgements}

The work of F.L.~was supported by the Swiss National Science Foundation (SNSF) under contract \href{https://data.snf.ch/grants/grant/211209}{TMSGI2\_211209}.
K.S.~is supported by the European Union under the Marie Sk{\l}odowska-Curie Actions (MSCA) Grant 101204018.
M.S.\ was supported by the Deutsche Forschungsgemeinschaft (DFG, German
Research Foundation) under grant 396021762 --- TRR 257 ``Particle Physics
Phenomenology after the Higgs Discovery''.
The Feynman diagrams have been drawn using \texttt{Jaxodraw}~\cite{Vermaseren:1994je,Binosi:2008ig}.

\appendix

\section{\label{app:distr}Distributions}
For the expansion into distributions we can follow Ref.~\cite{Alberti:2013kxa}
and use
\begin{align}
    \int_0^1  {\rm d} y \, \left[ \frac{1}{y^m} \right]_{+} f(y) &= \int_0^1  \frac{ {\rm d} y}{y^m} \left[ f(y) - \sum\limits_{i=0}^{m-1} \frac{f^{(i)}(0)}{i!} y^i \right]\,.
\end{align}
Specified to the
cases we encounter in our
calculations this leads to:
\begin{align}
    \int_0^1  {\rm d} y \, \frac{f(y)}{y^{1-n\epsilon}} &= \int_0^1  {\rm d} y \, \left[y^{-1+n\epsilon}\right]_{+} f(y) + \frac{1}{n \epsilon} f(0) \delta(y) ~,
    \nonumber\\
    \int_0^1  {\rm d} y \, \frac{f(y)}{y^{2-n\epsilon}} &= \int_0^1  {\rm d} y \, \left[y^{-2+n\epsilon}\right]_{+} f(y) + \left( \frac{1}{-1+n \epsilon} f(0) + \frac{1}{2n\epsilon} f^\prime(0) \right) \delta(y)  .
\end{align}
Note that
the terms in the parenthesis $\left[ \ldots \right]_{+}$ can be expanded in $\epsilon$.

While integrals over the full phase space involving plus distributions vanish,
\begin{align}
    \int\limits_{0}^{1} {\rm d}y \left[ \frac{\log^i(y)}{y} \right]_{+} &= 0 ~,
\end{align}
integrals with a lower cut on the photon energy give the finite contribution
\begin{align}
    \label{eq:plus-distr-ycut}
    \int\limits_{0}^{y_{\rm cut}} {\rm d}y \left[ \frac{\log^i(y)}{y} \right]_{+} &= -\int\limits_{y_{\rm cut}}^{1} {\rm d}y  \frac{\log^i(y)}{y} = \frac{\log^{i+1}(y_{\rm cut})}{i+1} ~,
\end{align}
which vanishes in the limit $y_{\rm cut} \to 1$.

\section{\label{app::formula}Analytic results for the regular part of the photon spectrum}

In this Appendix we show the analytic results of the regular parts of the spectrum introduced in Section~\ref{sub::spectrum_reg_sing}
and Eq.~(\ref{eq::spec}). The singular parts can be found in
Eqs.~(\ref{eq::spec_delta}) and~(\ref{eq::spec_plus}).
We express the amplitude in terms of $y=1-x$ and use the definition of iterated integrals introduced in Eq.~\eqref{eq::ints}, where we drop the argument $y$ to shorten the expressions.
At NLO and NNLO our results read
\begin{small}
\begin{align}
    Y_1^{\text{reg}} &=
    \CF
    \biggl\{
         \frac{6+3 y-2 y^2}{4}
        -\frac{(2-y) H_0}{2}
    \biggr\} \,,
\end{align}

\begin{align}
Y_2^{\text{reg}} &=
    \CF^2\biggl\{
    -\frac{32+56 y+413 y^2-148 y^3-8 y^4}{96 y}
    +\frac{H_0 P_{1}}{48 (1-y) y}
    +\frac{H_{-1,0} P_{2}}{12 (1-y) y^3}
    +\frac{H_{0,0} P_{3}}{12 (1-y)}
    \nonumber \\ &
    -\frac{H_{1,0} P_{4}}{12 (1-y) y^3}
    -\frac{\big(
        2+y-y^2-y^3\big) H_{-1,-1,0}}{2 y}
    +\frac{\big(
        2+y-y^2-y^3\big) H_{-1,0,0}}{y}
    \nonumber \\ &
    +\frac{\big(
        2+y-y^2-y^3\big) H_{-1,1,0}}{2 y}
    +\frac{\big(
        2+y-y^2-3 y^3\big) H_{0,-1,0}}{2 y}
    +\frac{3 \big(
        2-y+y^2-y^3\big) H_{0,0,0}}{2 (1-y)}
    \nonumber \\ &
    -\frac{
        (1-y) (2+y) H_{0,1,0}}{2}
    +\frac{\big(
        2+y-y^2-y^3\big) H_{1,-1,0}}{2 y}
    +\frac{\big(
        2+10 y-y^2\big) H_{1,0,0}}{4 y}
    \nonumber \\ &
    -\frac{\big(
        2+10 y-y^2\big) H_{1,1,0}}{4 y}
    - \frac{(9-5 y+y^2)\zeta_3}{4}
    + \biggl[
        \frac{P_{69}}{72 (1-y)}
        + \frac{ \big(2+y-y^2-y^3\big) H_{-1}}{24 y}
        \nonumber \\ &
        - \frac{ \big(9 + y^2 \big) H_{1}}{24}
        + \frac{1-y-y^2}{12} H_0
    \biggr] \pi^2
\biggr\}
\nonumber \\ &
+\CA \CF\biggl\{
     \frac{48+432 y+999 y^2-706 y^3-12 y^4}{288 y}
    -\frac{H_0 P_{5}}{144 (1-y) y}
    -\frac{H_{-1,0} P_{6}}{24 (1-y) y^3}
    \nonumber \\ &
    +\frac{\big(
        90-24 y+4 y^2+y^3-y^4\big) H_{0,0}}{24}
    +\frac{H_{1,0} P_{7}}{24 y^3}
    +\frac{\big(
        2+y-y^2-y^3\big) H_{-1,-1,0}}{4 y}
        \nonumber \\ &
    -\frac{\big(
        2+y-y^2-y^3\big) H_{-1,0,0}}{2 y}
    -\frac{\big(
        2+y-y^2-y^3\big) H_{-1,1,0}}{4 y}
    -\frac{\big(
        2+y-y^2-3 y^3\big) H_{0,-1,0}}{4 y}
    \nonumber \\ &
    -\frac{3 y^2 H_{0,0,0}}{4}
    +\frac{
        (1-y) (2+y) H_{0,1,0}}{4}
    -\frac{\big(
        2+y-y^2-y^3\big) H_{1,-1,0}}{4 y}
    +\frac{(9-5 y+y^2)\zeta_3}{8}
    \nonumber \\ &
    + \biggl[
        \frac{P_{68}}{144 (1-y)}
        - \frac{ \big(2+y-y^2-y^3\big) H_{-1}}{48 y}
        - \frac{ \big(2+y-y^2-y^3\big) H_{1}}{48 y}
        + \frac{3-y+y^2}{24} H_0
    \biggr] \pi^2
\biggr\}
\nonumber \\ &
+\CF \nl \TF\biggl\{
    -\frac{66+21 y-38 y^2}{72}
    +\frac{\big(
        38-33 y+7 y^2+6 y^3\big) H_0}{36 (1-y)}
    -\frac{(2-y) H_{0,0}}{2}
    \nonumber \\ &
    -\frac{\big(2+2 y-y^2\big) H_{1,0}}{6 y}
    + \frac{(y-2)\pi^2}{36}
\biggr\} \,,
\end{align}

\end{small}
and at N$^3$LO we have
\begin{small}
\begin{align}
Y_3^{\text{reg}} &=
    \CF \nl^2 \TF^2
    \biggl\{
     \frac{378-111 y-406 y^2}{648}
    -\frac{\big(
        298-225 y+65 y^2+114 y^3\big) H_0}{324 (1-y)}
    \nonumber \\ &
    +\frac{\big(
        26-27 y+17 y^2+2 y^3\big) H_{0,0}}{18 (1-y)}
    +\frac{\big(
        4+6 y-12 y^2+11 y^3\big) H_{1,0}}{27 (1-y) y}
    -\frac{-7 (-2+y) H_{0,0,0}}{9}
    \nonumber \\ &
    -\frac{2 \big(
        2+2 y-y^2\big) H_{0,1,0}}{9 y}
    -\frac{\big(
        2+2 y-y^2\big) H_{1,0,0}}{3 y}
    -\frac{\big(
        2+2 y-y^2\big) H_{1,1,0}}{9 y}
    \nonumber \\ &
    + \pi^2 \biggl[
        \frac{2-15y+37y^2-6y^3}{324(1-y)}
        - \frac{\big(2-y\big)H_{0}}{54}
        - \frac{\big(2+2y-y^2\big)H_{1}}{54y}
    \biggr]
    \biggr\}
    \nonumber \\ &
    +\CF \nh \nl \TF^2
    \biggl\{
     \frac{P_{8}}{2592 (1-y) y^3}
    -\frac{H_0 P_{9}}{81 (1-y) y^3}
    -\frac{\big(
        2+2 y-y^2\big) H_1}{32 y}
    +\frac{H_a P_{10}}{432 (1-y) y^2}
    \nonumber \\ &
    +\frac{H_{1,0} P_{11}}{27 (1-y) y^4}
    +\frac{H_{a,a} P_{12}}{216 (1-y) y^4}
    -\frac{\big(
        2+2 y-y^2\big) H_{b,a}}{24 y}
    -\frac{\big(
        2+2 y-y^2\big) H_{0,a,a}}{36 y}
    \nonumber \\ &
    -\frac{\big(
        2+2 y-y^2\big) H_{1,1,0}}{9 y}
    -\frac{\big(
        2+2 y-y^2\big) H_{1,a,a}}{12 y}
    \biggr\}
+\CF^2 \nl \TF\biggl\{
     \frac{P_{13}}{27648 (1-y)^2 y}
     \nonumber \\ &
    -\frac{H_0 P_{14}}{432 (1-y) y}
    -\frac{\big(
        3008-1659 y+422 y^2+36 y^3\big) H_1}{2304 y}
    -\frac{H_a P_{15}}{288 (1-y)^2 y}
    -\frac{H_{-1,0} P_{16}}{72 (1-y) y^3}
    \nonumber \\ &
    +\frac{H_{0,0} P_{17}}{144 (1-y) y}
    -\frac{3 \big(
        6+9 y-5 y^2\big) H_{0,1}}{64 y}
    +\frac{H_{1,0} P_{18}}{72 (1-y) (2-y) y^3}
    -\frac{H_{a,a} P_{19}}{288 (1-y)^2 (2-y) y^2}
    \nonumber \\ &
    -\frac{\big(
        41-78 y+33 y^2+y^3\big) H_{b,a}}{48 y}
    +\frac{H_{-1,-1,0} P_{20}}{18 y^3}
    +\frac{H_{-1,0,0} P_{21}}{36 (1-y) y^3}
    +\frac{H_{0,-1,0} P_{22}}{9 (1-y)^2 y^3}
    \nonumber \\ &
    -\frac{H_{0,0,0} P_{23}}{12 (1-y) (1+y)}
    -\frac{\big(
        10-5 y-10 y^2+8 y^4\big) H_{-1,1,0}}{9 (1-y) y}
    -\frac{H_{0,1,0} P_{24}}{18 (1-y)^2 y^3}
    -\frac{H_{0,a,a} P_{25}}{144 (1-y)^2 y}
    \nonumber \\ &
    -\frac{\big(
        6+9 y-5 y^2\big) H_{0,b,a}}{16 y}
    -\frac{\big(
        10-5 y-10 y^2+8 y^4\big) H_{1,-1,0}}{9 (1-y) y}
    -\frac{H_{1,0,0} P_{26}}{36 (1-y) y^3}
    \nonumber \\ &
    -\frac{\big(
        118-129 y+48 y^2+2 y^3\big) H_{1,a,a}}{48 y}
    +\frac{H_{2,1,0} P_{28}}{18 (1-y)^2 y}
    +\frac{H_{2,a,a} P_{29}}{36 (1-y)^2 y}
    \nonumber \\ &
    -\frac{\big(
        2+y-y^2-y^3\big) H_{-1,-1,-1,0}}{3 y}
    -\frac{5 \big(
        2+y-y^2-y^3\big) H_{-1,-1,0,0}}{6 y}
    -\frac{H_{1,1,0} P_{27}}{36 (1-y) y^3}
    \nonumber \\ &
    +\frac{\big(
        2-y^2-2 y^3\big) H_{-1,0,-1,0}}{3 y}
    +\frac{
        (2+y) \big(
                5+y-3 y^2\big) H_{-1,0,0,0}}{2 y}
    \nonumber \\ &
    +\frac{\big(
        2+3 y-y^2+y^3\big) H_{-1,0,1,0}}{6 y}
    +\frac{3 \big(
        2+y-y^2-y^3\big) H_{-1,1,0,0}}{2 y}
    \nonumber \\ &
    +\frac{\big(
        2+y-y^2-y^3\big) H_{-1,1,1,0}}{3 y}
    -\frac{\big(
        2+y-y^2+y^3\big) H_{0,-1,-1,0}}{3 y}
    \nonumber \\ &
    +\frac{\big(
        26+13 y-13 y^2-23 y^3\big) H_{0,-1,0,0}}{6 y}
    +\frac{2 \big(
        2+y-y^2-y^3\big) H_{0,-1,1,0}}{3 y}
    \nonumber \\ &
    +\frac{\big(
        4+3 y-2 y^2-3 y^3\big) H_{0,0,-1,0}}{3 y}
    +\frac{-5 \big(
        -2+y-y^2+y^3\big) H_{0,0,0,0}}{(1-y)}
    \nonumber \\ &
    -\frac{\big(
        4+9 y-20 y^2+7 y^3-y^4\big) H_{0,0,1,0}}{6 (1-y) y}
    -\frac{(9-5 y) H_{0,0,a,a}}{24}
    +\frac{2 \big(
        2+y-y^2-y^3\big) H_{0,1,-1,0}}{3 y}
    \nonumber \\ &
    -\frac{\big(
        2+9 y-23 y^2-2 y^3+11 y^4\big) H_{0,1,0,0}}{6 (1-y) y}
    -\frac{\big(
        22+25 y-14 y^2\big) H_{0,1,1,0}}{6 y}
    \nonumber \\ &
    -\frac{\big(
        6+9 y-5 y^2\big) H_{0,1,a,a}}{8 y}
    +\frac{3 \big(
        2+y-y^2-y^3\big) H_{1,-1,0,0}}{2 y}
    +\frac{\big(
        2+y-y^2-y^3\big) H_{1,-1,1,0}}{3 y}
    \nonumber \\ &
    +\frac{2 (1+y) \big(
        2-y^2\big) H_{1,0,-1,0}}{3 y}
    +\frac{\big(
        10+24 y-31 y^2+5 y^3-4 y^4\big) H_{1,0,0,0}}{4 (1-y) y}
    \nonumber \\ &
    -\frac{\big(
        4+15 y-2 y^2\big) H_{1,0,1,0}}{6 y}
    +\frac{\big(
        4+3 y-2 y^2\big) H_{1,0,a,a}}{24 y}
    +\frac{\big(
        2+y-y^2-y^3\big) H_{1,1,-1,0}}{3 y}
    \nonumber \\ &
    -\frac{\big(
        10+26 y-5 y^2\big) H_{1,1,0,0}}{6 y}
    -\frac{\big(
        2+10 y-y^2\big) H_{1,1,1,0}}{3 y}
    -\frac{\big(
        4+3 y-2 y^2\big) H_{1,2,1,0}}{3 y}
    \nonumber \\ &
    -\frac{\big(
        4+3 y-2 y^2\big) H_{1,2,a,a}}{6 y}
+\zeta_3 \biggl[
    \frac{P_{30}}{36 (1-y)}
    -\frac{
        (1-y) \big(
                4+5 y+3 y^2\big) H_{-1}}{6 y}
    \nonumber \\ &
    +\frac{\big(
        4-6 y+3 y^2-2 y^3\big) H_0}{6 (1-y)}
    -\frac{\big(
        2-17 y-y^2-2 y^3\big) H_1}{6 y}
 \biggr]
    +\frac{(129-37 y+63 y^2)\pi^4}{2160}
    \nonumber \\ &
    + \pi^2 \biggl[
          \frac{P_{73}}{864y^2(1-y)}
        - \frac{P_{72} H_{0}}{216(1-y^2)}
        - \frac{P_{70} H_{1}}{216y^3(1-y)}
        + \frac{P_{71} H_{-1}}{216y^3(1-y)}
        \nonumber \\ &
        - \frac{\big(2+y-y^2-y^3\big) H_{-1,-1}}{36y}
        + \frac{5\big(1+y^2\big) H_{-1,0}}{36}
        + \frac{\big(2+y-y^2-y^3\big) H_{-1,1}}{18y}
        \nonumber \\ &
        + \frac{\big(6+3y-3y^2-5y^3\big) H_{0,-1}}{36y}
        - \frac{\big(1+5y+y^2-4y^3\big) H_{0,0}}{12(1-y)}
        \nonumber \\ &
        - \frac{\big(12+14y-7y^2+2y^3\big) H_{0,1}}{36y}
        + \frac{\big(2+y-y^2-y^3\big) H_{1,-1}}{18y}
        \nonumber \\ &
        + \frac{\big(2-26y-y^2-4y^3\big) H_{1,0}}{72y}
        - \frac{\big(2+19y-y^2+y^3\big) H_{1,1}}{36 y}
    \biggr]
\biggr\}
\nonumber \\ &
+\CA \CF \nl \TF\biggl\{
     \frac{P_{31}}{165888 (1-y)^2 y}
    +\frac{H_0 P_{32}}{1296 (1-y) y}
    +\frac{\big(
        3008-1659 y+422 y^2+36 y^3\big) H_1}{4608 y}
    \nonumber \\ &
    +\frac{H_a P_{33}}{576 (1-y)^2 y}
    +\frac{H_{-1,0} P_{34}}{144 (1-y) y^3}
    -\frac{H_{0,0} P_{35}}{144 (1-y) y}
    +\frac{3 \big(
        6+9 y-5 y^2\big) H_{0,1}}{128 y}
    \nonumber \\ &
    -\frac{H_{1,0} P_{36}}{432 (1-y) (2-y) y^3}
    +\frac{H_{a,a} P_{37}}{576 (1-y)^2 (2-y) y^2}
    +\frac{\big(
        41-78 y+33 y^2+y^3\big) H_{b,a}}{96 y}
    \nonumber \\ &
    -\frac{H_{-1,-1,0} P_{38}}{36 y^3}
    -\frac{H_{-1,0,0} P_{39}}{72 (1-y) y^3}
    +\frac{\big(
        10-5 y-10 y^2+8 y^4\big) H_{-1,1,0}}{18 (1-y) y}
    -\frac{H_{0,-1,0} P_{40}}{18 (1-y)^2 y^3}
    \nonumber \\ &
    +\frac{H_{0,0,0} P_{41}}{72 (1-y) (1+y)}
    +\frac{H_{0,1,0} P_{42}}{72 (1-y)^2 y^3}
    +\frac{H_{0,a,a} P_{43}}{288 (1-y)^2 y}
    +\frac{\big(
        6+9 y-5 y^2\big) H_{0,b,a}}{32 y}
    \nonumber \\ &
    +\frac{\big(
        10-5 y-10 y^2+8 y^4\big) H_{1,-1,0}}{18 (1-y) y}
    +\frac{H_{1,0,0} P_{44}}{72 y^3}
    +\frac{H_{1,1,0} P_{45}}{72 y^3}
    -\frac{H_{2,1,0} P_{46}}{36 (1-y)^2 y}
    \nonumber \\ &
    +\frac{\big(
        118-129 y+48 y^2+2 y^3\big) H_{1,a,a}}{96 y}
    +\frac{\big(
        2+y-y^2-y^3\big) H_{-1,-1,-1,0}}{6 y}
    \nonumber \\ &
    +\frac{5 \big(
        2+y-y^2-y^3\big) H_{-1,-1,0,0}}{12 y}
    -\frac{\big(
        2-y^2-2 y^3\big) H_{-1,0,-1,0}}{6 y}
    -\frac{H_{2,a,a} P_{47}}{72 (1-y)^2 y}
    \nonumber \\ &
    -\frac{
        (2+y) \big(
                5+y-3 y^2\big) H_{-1,0,0,0}}{4 y}
    -\frac{\big(
        2+3 y-y^2+y^3\big) H_{-1,0,1,0}}{12 y}
    \nonumber \\ &
    -\frac{3 \big(
        2+y-y^2-y^3\big) H_{-1,1,0,0}}{4 y}
    -\frac{\big(
        2+y-y^2-y^3\big) H_{-1,1,1,0}}{6 y}
    -\frac{(6+7 y+21 y^2)\pi^4}{1440}
    \nonumber \\ &
    +\frac{\big(
        2+y-y^2+y^3\big) H_{0,-1,-1,0}}{6 y}
    -\frac{\big(
        26+13 y-13 y^2-23 y^3\big) H_{0,-1,0,0}}{12 y}
    \nonumber \\ &
    -\frac{\big(
        2+y-y^2-y^3\big) H_{0,-1,1,0}}{3 y}
    -\frac{\big(
        4+3 y-2 y^2-3 y^3\big) H_{0,0,-1,0}}{6 y}
    -\frac{5 y^2 H_{0,0,0,0}}{2}
    \nonumber \\ &
    +\frac{\big(
        6+15 y-6 y^2+y^3\big) H_{0,0,1,0}}{12 y}
    +\frac{(9-5 y) H_{0,0,a,a}}{48}
    -\frac{\big(
        2+y-y^2-y^3\big) H_{0,1,-1,0}}{3 y}
    \nonumber \\ &
    +\frac{\big(
        2+12 y-6 y^2-11 y^3\big) H_{0,1,0,0}}{12 y}
    +\frac{\big(
        9+11 y-6 y^2\big) H_{0,1,1,0}}{6 y}
    +\frac{\big(
        6+9 y-5 y^2\big) H_{0,1,a,a}}{16 y}
    \nonumber \\ &
    -\frac{3 \big(
        2+y-y^2-y^3\big) H_{1,-1,0,0}}{4 y}
    -\frac{\big(
        2+y-y^2-y^3\big) H_{1,-1,1,0}}{6 y}
    -\frac{
        (1+y) \big(
                2-y^2\big) H_{1,0,-1,0}}{3 y}
    \nonumber \\ &
    +\frac{\big(
        1-2 y^2\big) H_{1,0,0,0}}{4}
    +\frac{H_{1,0,1,0}}{4}
    -\frac{\big(
        4+3 y-2 y^2\big) H_{1,0,a,a}}{48 y}
    -\frac{\big(
        2+y-y^2-y^3\big) H_{1,1,-1,0}}{6 y}
    \nonumber \\ &
    +\frac{\big(
        4+3 y-2 y^2\big) H_{1,2,1,0}}{6 y}
    +\frac{\big(
        4+3 y-2 y^2\big) H_{1,2,a,a}}{12 y}
+\zeta_3 \biggl[
    -\frac{P_{48}}{72 (1-y)}
    \nonumber \\ &
    +\frac{
        (1-y) \big(
                4+5 y+3 y^2\big) H_{-1}}{12 y}
    -\frac{\big(
        4+y+2 y^2\big) H_0}{12}
    +\frac{\big(
        8+5 y-4 y^2-2 y^3\big) H_1}{12 y}
 \biggr]
 \nonumber \\ &
 + \pi^2 \biggl[
        - \frac{P_{74}}{2592y^2(1-y)}
        - \frac{P_{75} H_{0}}{432(1-y^2)}
        + \frac{P_{76} H_{1}}{432y^3(1-y)}
        - \frac{P_{77} H_{-1}}{432y^3(1-y)}
        \nonumber \\ &
        + \frac{\big(2+y-y^2-y^3\big) H_{-1,-1}}{72y}
        - \frac{5\big(1+y^2\big) H_{-1,0}}{72}
        - \frac{\big(2+y-y^2-y^3\big) H_{-1,1}}{36y}
        \nonumber \\ &
        - \frac{\big(6+3y-3y^2-5y^3\big) H_{0,-1}}{72y}
        + \frac{\big(5-2y+4y^2\big) H_{0,0}}{24}
        + \frac{\big(8+11y-5y^2+2y^3\big) H_{0,1}}{72y}
        \nonumber \\ &
        - \frac{\big(2+y-y^2-y^3\big) H_{1,-1}}{36y}
        + \frac{\big(1+y^2\big) H_{1,0}}{36}
        - \frac{\big(2+y-y^2-y^3\big) H_{1,1}}{72 y}
    \biggr]
\biggr\}
\nonumber \\ &
+ \Nc^3\biggl\{
    -\frac{18144-98460 y-57567 y^2+625304 y^3}{165888 y}
    +\frac{H_0 P_{49}}{82944 (1-y) (2-y) y}
    \nonumber \\ &
    +\frac{H_{0,0} P_{50}}{4608 (1-y) (2-y)^3}
    -\frac{H_{1,0} P_{51}}{3456 (1-y) (2-y)^3 y^2}
    -\frac{H_{0,0,0} P_{52}}{2304 (1-y) (2-y)^4}
    \nonumber \\ &
    -\frac{H_{0,1,0} P_{53}}{2304 (1-y) y}
    -\frac{H_{1,0,0} P_{54}}{2304 (1-y) (2-y)^4 y}
    -\frac{H_{1,1,0} P_{55}}{2304 (1-y) (2-y)^4 y}
    -\frac{H_{0,0,0,0} P_{56}}{128 (1-y)^2}
    \nonumber \\ &
    -\frac{H_{0,0,1,0} P_{57}}{384 (1-y) y}
    -\frac{H_{0,1,0,0} P_{58}}{384 (1-y) y}
    +\frac{H_{0,1,1,0} P_{59}}{384 y}
    -\frac{H_{1,0,0,0} P_{60}}{128 (1-y) (2-y)^5 y}
    +\frac{H_{1,0,1,0} P_{61}}{384 (1-y) y}
    \nonumber \\ &
    +\frac{H_{1,1,0,0} P_{62}}{384 (1-y) (2-y)^5 y}
    +\frac{H_{1,1,1,0} P_{63}}{384 (1-y) (2-y)^5 y^2}
    -\frac{\big(
        60-11 y-6 y^2\big) H_{0,0,0,0,0}}{32 (1-y)}
    \nonumber \\ &
    -\frac{\big(
        8-13 y^2+7 y^3\big) H_{0,0,0,1,0}}{32 (1-y) y}
    -\frac{\big(
        10+2 y-15 y^2+6 y^3\big) H_{0,0,1,0,0}}{32 (1-y) y}
    \nonumber \\ &
    +\frac{\big(
        2-2 y-y^2+2 y^3\big) H_{0,0,1,1,0}}{32 (1-y) y}
    -\frac{\big(
        12-19 y+13 y^2+2 y^3\big) H_{0,1,0,0,0}}{32 (1-y) y}
    \nonumber \\ &
    -\frac{\big(
        6-7 y+3 y^2+y^3\big) H_{0,1,0,1,0}}{32 (1-y) y}
    -\frac{\big(
        6+21 y-25 y^2-y^3\big) H_{0,1,1,0,0}}{32 (1-y) y}
    +\frac{(6+y) H_{2,1,1,1,0}}{16}
    \nonumber \\ &
    +\frac{\big(
        2-28 y-11 y^2\big) H_{0,1,1,1,0}}{32 y}
    -\frac{\big(
        15+13 y-12 y^2-6 y^3\big) H_{1,0,0,0,0}}{32 (1-y) y}
    -\frac{(3+2 y) H_{1,0,1,1,0}}{32}
    \nonumber \\ &
    -\frac{\big(
        4+16 y-26 y^2+5 y^3\big) H_{1,0,0,1,0}}{32 (1-y) y}
    -\frac{\big(
        4+23 y-37 y^2+7 y^3\big) H_{1,0,1,0,0}}{32 (1-y) y}
    \nonumber \\ &
    +\frac{\big(
        18+30 y-31 y^2-5 y^3\big) H_{1,1,0,0,0}}{64 (1-y) y}
    -\frac{\big(
        2+10 y-y^2\big) H_{1,1,0,1,0}}{32 y}
    -\frac{(6+y) H_{2,1,0,0,0}}{8}
    \nonumber \\ &
    -\frac{\big(
        2+5 y-y^2\big) H_{1,1,1,0,0}}{8 y}
    -\frac{\big(
        2+30 y-y^2\big) H_{1,1,1,1,0}}{32 y}
    +\frac{(6+y) H_{2,1,1,0,0}}{16}
    \nonumber \\ &
    -\frac{(296-221 y-51 y^2)\zeta_5}{128 (1-y)}
    +\frac{(270-253 y-33 y^2)\pi^2 \zeta_3}{384 (1-y)}
+\zeta_3 \biggl[
     \frac{P_{64}}{384 (1-y) (2-y)^4 y}
     \nonumber \\ &
    +\frac{H_0 P_{65}}{192 (1-y)}
    -\frac{H_1 P_{66}}{192 (1-y) (2-y)^5 y^2}
    -\frac{\big(
        2-6 y+5 y^2\big) H_{0,0}}{16 (1-y)}
    -\frac{(6+y) H_{2,1}}{8}
    \nonumber \\ &
    -\frac{\big(
        8-65 y+48 y^2+14 y^3\big) H_{0,1}}{32 (1-y) y}
    -\frac{(2-y) \big(
        2+4 y-5 y^2\big) H_{1,0}}{32 (1-y) y}
    +\frac{3 \big(
        2+10 y-y^2\big) H_{1,1}}{32 y}
 \biggr]
 \nonumber \\ &
+\pi^4 \biggl[
    -\frac{P_{67}}{138240 (1-y)}
    -\frac{\big(
        192-293 y+106 y^2\big) H_0}{11520 (1-y)}
    +\frac{\big(
        104+375 y-58 y^2\big) H_1}{11520 y}
 \biggr]
 \nonumber \\ &
 + \pi^2 \biggl[
      \frac{P_{78}}{82944 y (1-y) (2-y)^3}
    - \frac{P_{79} H_{0}}{4608 y (1-y)}
    - \frac{P_{80} H_{1}}{13824 y (1-y) (2-y)^4}
    \nonumber \\ &
    + \frac{P_{81} H_{0,0}}{2304 (1-y)^2}
    + \frac{P_{82} H_{0,1}}{2304y}
    - \frac{P_{83} H_{1,0}}{2304 y^2 (1-y)}
    + \frac{P_{84} H_{1,1}}{2304 y^2 (1-y) (2-y)^5}
    \nonumber \\ &
    - \frac{\big(24-25y+9y^2\big) H_{0,0,0}}{192 (1-y)}
    + \frac{\big(2-2y-y^2+2y^3\big) H_{0,0,1}}{192 y (1-y)}
    - \frac{\big(2-31y-9y^2\big) H_{0,1,0}}{192y}
    \nonumber \\ &
    + \frac{\big(2-28y-11y^2\big) H_{0,1,1}}{192 y}
    - \frac{\big(2+38y-60y^2+13y^3\big) H_{1,0,0}}{192 y (1-y)}
    - \frac{\big(3+2y\big) H_{1,0,1}}{192}
    \nonumber \\ &
    + \frac{\big(14+50y-11y^2\big) H_{1,1,0}}{384 y}
    - \frac{\big(2+30y-y^2\big) H_{1,1,1}}{192 y}
    + \frac{\big(6+y\big) H_{2,1,1}}{96}
 \biggr]
\biggr\}\,.
\end{align}

\end{small}
For convenience, we introduced the polynomials
\begin{small}
\begin{align}
    P_1 &= 32-24 y-23 y^2-39 y^3+22 y^4-4 y^5, 
     \nonumber \\ 
P_2 &= 8-32 y+45 y^3-10 y^4-11 y^5-5 y^6-2 y^7+y^8, 
     \nonumber \\ 
P_3 &= 6-27 y+14 y^2-3 y^3+2 y^4-y^5, 
     \nonumber \\ 
P_4 &= 8-32 y+8 y^2+47 y^3-46 y^4-2 y^5+8 y^6, 
     \nonumber \\ 
P_5 &= 48+412 y-591 y^2+128 y^3+81 y^4-6 y^5, 
     \nonumber \\ 
P_6 &= 8-32 y+45 y^3-10 y^4-11 y^5-5 y^6-2 y^7+y^8, 
     \nonumber \\ 
P_7 &= 8-24 y+6 y^2+44 y^3-17 y^4-2 y^5, 
     \nonumber \\ 
P_8 &= 768-3264 y+4640 y^2+610 y^3-14070 y^4+23584 y^5-16792 y^6+5265 y^7-450 y^8
\nonumber \nonumber \\ &
-57 y^9+9 y^{10}, 
     \nonumber \\ 
P_9 &= 24-96 y+134 y^2+114 y^3-189 y^4+76 y^5, 
     \nonumber \\ 
P_{10} &= \sqrt{4-y}\sqrt{y} \big(
        232+306 y-1279 y^2+1085 y^3-297 y^4+y^5+6 y^6\big), 
     \nonumber \\ 
P_{11} &= 8-36 y+60 y^2+18 y^3-84 y^4+54 y^5-11 y^6, 
     \nonumber \\ 
P_{12} &= 32-126 y+186 y^2+167 y^3-381 y^4+201 y^5-19 y^6-6 y^7, 
     \nonumber \\ 
P_{13} &= 12288-47328 y+524580 y^2-1178622 y^3+970722 y^4-316577 y^5+49742 y^6
\nonumber \nonumber \\ &
-9103 y^7+946 y^8-48 y^9, 
     \nonumber \\ 
P_{14} &= 576-713 y+1620 y^2-2812 y^3+783 y^4-66 y^5, 
     \nonumber \\ 
P_{15} &= \sqrt{4-y}\sqrt{y} \big(
        384-1408 y+1852 y^2-1025 y^3+265 y^4-34 y^5+2 y^6\big), 
     \nonumber \\ 
P_{16} &= 64-144 y-48 y^2+59 y^3-13 y^4+88 y^5-104 y^6-27 y^7+13 y^8, 
     \nonumber \\ 
P_{17} &= 160-484 y+301 y^2-435 y^3+62 y^4-74 y^5+26 y^6, 
     \nonumber \\ 
P_{18} &= 128-256 y-74 y^2+533 y^3-444 y^4-286 y^5+395 y^6-102 y^7+4 y^8, 
     \nonumber \\ 
P_{19} &= 192-108 y-1430 y^2+2460 y^3-1651 y^4+602 y^5-105 y^6+4 y^7, 
     \nonumber \\ 
P_{20} &= 8-24 y-4 y^2+31 y^3+y^4-10 y^5+y^6-y^7, 
     \nonumber \\ 
P_{21} &= 40-160 y-80 y^2+265 y^3+30 y^4-55 y^5-89 y^6-10 y^7+5 y^8, 
     \nonumber \\ 
P_{22} &= 8-40 y+28 y^2+23 y^3-26 y^4+41 y^5-52 y^6+24 y^7, 
     \nonumber \\ 
P_{23} &= 22+105 y+43 y^2-33 y^3-31 y^4-3 y^5+3 y^6, 
     \nonumber \\ 
P_{24} &= 16-80 y+82 y^2-31 y^3+39 y^4-26 y^5+13 y^6+y^7-3 y^8+y^9, 
     \nonumber \\ 
P_{25} &= 150-336 y+267 y^2-86 y^3+19 y^4-2 y^5, 
     \nonumber \\ 
P_{26} &= 40-160 y-35 y^2+261 y^3-246 y^4+18 y^5+34 y^6-4 y^7+2 y^8, 
     \nonumber \\ 
P_{27} &= 40-160 y+320 y^2-347 y^3+104 y^4-3 y^5+8 y^6+4 y^7-2 y^8, 
     \nonumber \\ 
P_{28} &= 38-136 y+209 y^2-118 y^3+17 y^4+2 y^5, 
     \nonumber \\ 
P_{29} &= 38-136 y+209 y^2-118 y^3+17 y^4+2 y^5, 
     \nonumber \\ 
P_{30} &= 69-320 y+223 y^2+12 y^3+8 y^4-4 y^5, 
     \nonumber \\ 
P_{31} &= 4608-448416 y-659436 y^2+4185274 y^3-4799654 y^4+1822435 y^5-149226 y^6
 \nonumber\nonumber \\ &
+27309 y^7-2838 y^8+144 y^9, 
     \nonumber \\ 
P_{32} &= 540+8440 y-7893 y^2+284 y^3+4011 y^4-99 y^5, 
     \nonumber \\ 
P_{33} &=  \sqrt{4-y}\sqrt{y} \big(
        384-1408 y+1852 y^2-1025 y^3+265 y^4-34 y^5+2 y^6\big), 
     \nonumber \\ 
P_{34} &= 64-144 y-48 y^2+59 y^3-13 y^4+88 y^5-104 y^6-27 y^7+13 y^8, 
     \nonumber \\ 
P_{35} &= 80+1348 y-1952 y^2+1063 y^3+7 y^4-37 y^5+13 y^6, 
     \nonumber \\ 
P_{36} &= 384-984 y+1718 y^2+749 y^3-4359 y^4+4207 y^5-1163 y^6-24 y^7+12 y^8, 
     \nonumber \\ 
P_{37} &= 192-108 y-1430 y^2+2460 y^3-1651 y^4+602 y^5-105 y^6+4 y^7, 
     \nonumber \\ 
P_{38} &= 8-24 y-4 y^2+31 y^3+y^4-10 y^5+y^6-y^7, 
     \nonumber \\ 
P_{39} &= 40-160 y-80 y^2+265 y^3+30 y^4-55 y^5-89 y^6-10 y^7+5 y^8, 
     \nonumber \\ 
P_{40} &= 8-40 y+28 y^2+23 y^3-26 y^4+41 y^5-52 y^6+24 y^7, 
     \nonumber \\ 
P_{41} &= 736-161 y-634 y^2+236 y^3-141 y^4-9 y^5+9 y^6, 
     \nonumber \\ 
P_{42} &= 32-160 y+409 y^2-412 y^3-3 y^4+240 y^5-104 y^6+26 y^7-6 y^8+2 y^9, 
     \nonumber \\ 
P_{43} &= 150-336 y+267 y^2-86 y^3+19 y^4-2 y^5, 
     \nonumber \\ 
P_{44} &= 40-120 y+155 y^2+336 y^3-139 y^4-2 y^5+2 y^6-2 y^7, 
     \nonumber \\ 
P_{45} &= 40-120 y+287 y^2-52 y^3-25 y^4-4 y^5-2 y^6+2 y^7, 
     \nonumber \\ 
P_{46} &= 38-136 y+209 y^2-118 y^3+17 y^4+2 y^5, 
     \nonumber \\ 
P_{47} &= 38-136 y+209 y^2-118 y^3+17 y^4+2 y^5, 
     \nonumber \\ 
P_{48} &= 60-227 y+199 y^2+6 y^3+8 y^4-4 y^5, 
     \nonumber \\ 
P_{49} &= 18144-717664 y+1393886 y^2-1133657 y^3+67075 y^4+119280 y^5, 
     \nonumber \\ 
P_{50} &= 274192-790576 y+1022744 y^2-729184 y^3+291089 y^4-60673 y^5+5656 y^6-168 y^7, 
     \nonumber \\ 
P_{51} &= 3024-61016 y+183708 y^2-236394 y^3+76836 y^4+100086 y^5-104001 y^6
 \nonumber\nonumber \\ &
+36935 y^7-5091 y^8+126 y^9, 
     \nonumber \\ 
P_{52} &= 272992-1044080 y+1562576 y^2-1232952 y^3+571342 y^4-159511 y^5+24578 y^6
 \nonumber\nonumber \\ &
-1221 y^7-126 y^8, 
     \nonumber \\ 
P_{53} &= 6314-2172 y-4044 y^2+2395 y^3-1047 y^4-42 y^5, 
     \nonumber \\ 
P_{54} &= 97760-171536 y+8736 y^2+228808 y^3-283210 y^4+175611 y^5-62971 y^6+12211 y^7
 \nonumber\nonumber \\ &
-831 y^8-42 y^9, 
     \nonumber \\ 
P_{55} &= 32096-37872 y-79008 y^2+208376 y^3-194594 y^4+92685 y^5-23367 y^6+3151 y^7
 \nonumber\nonumber \\ &
-321 y^8+42 y^9, 
     \nonumber \\ 
P_{56} &= 1180-1728 y+726 y^2-110 y^3+7 y^4+12 y^5-7 y^6, 
     \nonumber \\ 
P_{57} &= 256-12 y-350 y^2+144 y^3+4 y^4-5 y^5+7 y^6, 
     \nonumber \\ 
P_{58} &= 210-40 y-230 y^2+66 y^3+4 y^4-5 y^5+7 y^6, 
     \nonumber \\ 
P_{59} &= 218+102 y-78 y^2+18 y^3-2 y^4-7 y^5, 
     \nonumber \\ 
P_{60} &= 10368-5472 y-57616 y^2+122720 y^3-112960 y^4+57278 y^5-16673 y^6+2248 y^7
 \nonumber\nonumber \\ &
+375 y^8-302 y^9+75 y^{10}-7 y^{11}, 
     \nonumber \\ 
P_{61} &= 104+233 y-303 y^2+8 y^3-4 y^4+5 y^5-7 y^6, 
     \nonumber \\ 
P_{62} &= 13696-6208 y-70768 y^2+138280 y^3-113064 y^4+47190 y^5-10067 y^6+1549 y^7
 \nonumber\nonumber \\ &
-730 y^8+334 y^9-75 y^{10}+7 y^{11}, 
     \nonumber \\ 
P_{63} &= 3520-26656 y+95616 y^2-179744 y^3+192708 y^4-125322 y^5+52036 y^6
 \nonumber\nonumber \\ &
-14721 y^7+2727 y^8+66 y^9-262 y^{10}+75 y^{11}-7 y^{12}, 
     \nonumber \\ 
P_{64} &= 1760-10016 y+16128 y^2+2464 y^3-25794 y^4+24018 y^5-9947 y^6+2078 y^7
 \nonumber\nonumber \\ &
-219 y^8+14 y^9, 
     \nonumber \\ 
P_{65} &= 54-71 y-27 y^2+8 y^3+5 y^4-7 y^5, 
     \nonumber \\ 
P_{66} &= 1760-9808 y+45568 y^2-105240 y^3+126518 y^4-85905 y^5+34147 y^6-7867 y^7
 \nonumber\nonumber \\ &
+635 y^8+476 y^9-298 y^{10}+75 y^{11}-7 y^{12}, 
     \nonumber \\ 
P_{67} &= 174-2916 y+6144 y^2-2116 y^3-25 y^4+35 y^5,
\nonumber \\ 
P_{68} &= 8 - 28 y + 17 y^2 + 5 y^3 + 2 y^4 - y^5,
\nonumber \\
P_{69} &= -31 + 31 y + 30 y^2 - 23 y^3 - 2 y^4 + y^5,
\nonumber \\ 
P_{70} &= 40 - 160 y + 234 y^2 - 113 y^3 - 108 y^4 + 49 y^5 + 26 y^6 + 4 y^7 - 
 2 y^8,
\nonumber \\
P_{71} &= 8 - 32 y - 20 y^2 + 55 y^3 + 10 y^4 - 11 y^5 - 21 y^6 - 2 y^7 + y^8,
\nonumber \\
P_{72} &= 132 + 73 y - 299 y^2 - 43 y^3 + 107 y^4 + 3 y^5 - 3 y^6,
\nonumber \\ 
P_{73} &= 128 - 416 y + 1406 y^2 - 859 y^3 - 1053 y^4 + 522 y^5 + 50 y^6 - 
 26 y^7,
 \nonumber \\
P_{74} &= 192 - 624 y + 88 y^2 - 936 y^3 + 1991 y^4 - 273 y^5 + 75 y^6 - 39 y^7,
\nonumber \\
P_{75} &= 136 - 75 y - 20 y^2 + 54 y^3 - 47 y^4 - 3 y^5 + 3 y^6,
\nonumber \\
P_{76} &= 40 - 160 y + 321 y^2 - 105 y^3 - 185 y^4 + 73 y^5 + 20 y^6 + 4 y^7 - 
 2 y^8,
\nonumber \\
P_{77} &= 8 - 32 y - 20 y^2 + 55 y^3 + 10 y^4 - 11 y^5 - 21 y^6 - 2 y^7 + y^8,
\nonumber \\
P_{78} &=-25344 - 1042816 y + 2510712 y^2 - 1375772 y^3 - 788722 y^4 + 1105035 y^5 
\nonumber \\ &
- 413921 y^6 + 54480 y^7 - 504 y^8,
\nonumber \\
P_{79} &= -176 - 4132 y + 6149 y^2 - 848 y^3 - 925 y^4 - 14 y^5,
\nonumber \\
P_{80} &= 32096 - 37872 y - 79008 y^2 + 208376 y^3 - 194594 y^4 + 92685 y^5 - 23367 y^6 
\nonumber \\ &
+ 3151 y^7 - 321 y^8 + 42 y^9,
\nonumber \\
P_{81} &= -732 + 3118 y - 3086 y^2 + 840 y^3 - 39 y^4 - 12 y^5 + 7 y^6,
\nonumber \\
P_{82} &= 218 + 102 y - 78 y^2 + 18 y^3 - 2 y^4 - 7 y^5,
\nonumber \\
P_{83} &= 88 - 590 y - 73 y^2 + 760 y^3 - 41 y^4 - 92 y^5 - 5 y^6 + 7 y^7,
\nonumber \\
P_{84} &= 3520 - 26656 y + 95616 y^2 - 179744 y^3 + 192708 y^4 - 125322 y^5 + 52036 y^6 
\nonumber \\ &
- 14721 y^7 + 2727 y^8 + 66 y^9 - 262 y^{10} + 75 y^{11} - 7 y^{12}.
\end{align}
\end{small}

\bibliographystyle{jhep}
\bibliography{main.bib}

\end{document}